\documentclass[aps,prb,preprint, showpacs]{revtex4}

\usepackage{epsfig}
\usepackage{graphicx}

\def\be{\begin{equation}}
\def\ee{\end{equation}}
\def\bea{\begin{eqnarray}}
\def\eea{\end{eqnarray}}

\begin{document}

\title{Theory of Electric Dipole Spin Resonance in a Parabolic Quantum Well}
\author{Al. L. Efros}
\affiliation{Naval Research Laboratory, Washington, DC 20375, USA}
\author{E. I. Rashba }
\affiliation{Department of Physics, Harvard University, Cambridge, Massachusetts 02138, USA}
\begin{abstract}
A theory of Electric Dipole Spin Resonance (EDSR), that is caused by various mechanisms of spin-orbit coupling, is developed as applied to free electrons in a parabolic quantum well. Choosing a parabolic shape of the well has allowed us to find explicit expressions for the EDSR intensity and its dependence on the magnetic field direction in terms of the  basic parameters of the Hamiltonian. By using these expressions, we have investigated and compared the effect of specific mechanisms of spin orbit (SO) coupling and different polarizations of ac electric field on the intensity of EDSR. It is our basic assumption  that the SO coupling energy is small compared with all different competing energies (the confinement energy, and the cyclotron and Zeeman energies) that allowed us to describe all SO coupling mechanisms in the framework of the same general approach. For this purpose, we have developed an operator formalism for calculating matrix elements of the transitions between different quantum levels. To make these calculations efficient enough and to derive explicit and concise expressions for the EDSR intensity, we have established a set of remarkable identities relating the eigenfrequencies and the angles defining the spatial orientation of the quantizing magnetic field $\mbox{\boldmath$B$}(\theta,\phi)$. Applicability of these identities is not restricted by EDSR and we expect them to be useful for the general theory of parabolic quantum wells. The angular dependences of the EDSR intensity, found for various SO coupling mechanisms, show a fine structure consisting of alternating up- and down-cusps originating from repopulating different quantum levels and their spin sublevels. Angular dependences of the EDSR intensity are indicative of the relative contributions of the competing mechanisms of SO coupling. Our results show that electrical manipulating electron spins in quantum wells is generally highly efficient, especially by an in-plane ac electric field. 
\end{abstract}

\date{\today, SashaLongFinal.tex}

\maketitle

\section{Introduction}

Efficient manipulation of electron spins by an external ac field is one of the central problems of semiconductor spintronics,\cite{Wolf,ZFDS04} quantum computing\cite{LDV} and information processing.\cite{ALS02} The original proposals regarding spin manipulation were based on using a time dependent magnetic field $\tilde{H}(t)$. However, there is a growing understanding of the advantages of  spin manipulation by a time-dependent electric field $\tilde{E}(t)$ that allows the access to electron spins with nanometer precision and can provide much stronger coupling to electron spins through various mechanisms of SO interaction.\cite{RS61,RS91} Different options of electrical manipulating electron spins in semiconductor nanostructures range from adiabatic pumping spin currents from quantum dots\cite{WPMU03} to manipulating electron spins in quantum wells (QW) at the spin resonance frequency by employing various mechanisms of SO coupling. Spin orbit interaction
\be
{\hat H}_{SO}={\hat H}_{\rm orb}({\hat{\mbox{\boldmath$k$}}},\mbox{\boldmath$\sigma$})+{\hat H}_Z(\mbox{\boldmath$r$},\mbox{\boldmath$\sigma$})
\label{eq0}
\ee
can be usually represented as a sum of the orbital contribution ${\hat H}_{\rm orb}({\hat{\mbox{\boldmath$k$}}},\mbox{\boldmath$\sigma$})$ depending on the momenta ${\hat{\mbox{\boldmath$k$}}}$ and Pauli matrices $\mbox{\boldmath$\sigma$}$, and the Zeeman contribution ${\hat H}_Z(\mbox{\boldmath$r$},\mbox{\boldmath$\sigma$})$ depending on the coordinates $\mbox{\boldmath$r$}$ and matrices $\mbox{\boldmath$\sigma$}$. Because both terms include orbital operators ($\hat{\mbox{\boldmath$k$}}$ or $\mbox{\boldmath$r$}$) and matrices $\mbox{\boldmath$\sigma$}$, they represent different mechanisms of SO coupling.

For two-dimensional (2D) electrons in QWs, two basic mechanisms of the orbital SO coupling are directly related to the QW symmetry properties. They stem from the structure inversion asymmetry (SIA) mechanism described by the Rashba term, \cite{R60,BR84} and  the $A_3B_5$ compound bulk inversion asymmetry (BIA) mechanism described by the Dresselhaus term.\cite{D55} In the principle crystal axes, the bulk Dresselhaus 3D spin-orbit interaction $\hat{\cal H}_D$ can be written as\cite{RS61}
\be
\hat{\cal H}_D=\delta(\mbox{\boldmath $\sigma$}\cdot\hat{\mbox{\boldmath $\kappa$}}),~~{\rm where}~~
{\hat\kappa}_x={\hat k}_y{\hat k}_x{\hat k}_y-{\hat k}_z{\hat k}_x{\hat k}_z~,
\label{eq11}
\ee 
${\hat\kappa}_y$ and ${\hat\kappa}_z$ can be derived from ${\hat\kappa}_x$ by cyclic permutations, and $\delta$ is a parameter. Here ${\hat k}_j$ ($j=x,y,z$) are the projections of the momentum operator $\hat{\mbox{\boldmath $k$}}=-i\mbox{\boldmath $\nabla$}+e\mbox{\boldmath $A$}/\hbar c$ of an electron, $\mbox{\boldmath $A$}$ is the vector-potential of the magnetic field $\mbox{\boldmath $B$}(\theta,\phi)$, $\theta$ and $\phi$ are the polar angle and the azimuth of $\mbox{\boldmath $B$}$, and $-e$ is the electron charge. In a strong confinement limit, when carriers are in a quasi-2D regime,  $\hat{\cal H}_D$ reduces to a 2D Dresselhaus Hamiltonian. For a rectangular [001] QW of the width $d$,
\be
{\hat H}_D=\alpha_D(\sigma_x\hat{k}_x-\sigma_y\hat{k}_y)
\label{eqDr}
\ee
 with $\alpha_D=-\delta\langle k_z^2\rangle=-\delta(\pi/d)^2$.\cite{LMR85,BR85,DK86,RS88,PP95}
The 2D Dresselhaus Hamiltonian possesses a discrete symmetry $\mbox{\boldmath $C$}_{2v}$ of the [001] face of a cubic lattice of the $\mbox{\boldmath $T$}_d$ symmetry typical of the zinc blende modification of A$_3$B$_5$ crystals, while the Rashba SO Hamiltonian
\be
{\hat H}_R=\alpha_R(\sigma_x\hat{k}_y-\sigma_y\hat{k}_x)
\label{eqR}
\ee
possesses a continuous group $\mbox{\boldmath $C$}_{\infty v}$ of the rotations about the [001] axis.  Experimental data of Refs.~\onlinecite{Jus95,Knap96,Mil03} suggest that for GaAs QWs both SIA and BIA terms  are of the same order of magnitude.  It is a general consensus that in narrow-gap compounds the SIA mechanism typically dominates,\cite{Koga02} however, the ratio of the coupling constants of only $\alpha_R/\alpha_D\approx 2$ has been recently reported for InAs QWs of width $d=15$ nm.\cite{Gan04} The tunability of $\alpha_R$ by gate voltage\cite{NATE97,ELSL97,G00} is believed to be of critical importance for the operation of semiconductor spintronic devices.

The spatial dependence of Zeeman energy ${\hat H}_Z(\mbox{\boldmath$r$},\mbox{\boldmath$\sigma$})$ stems either from the inhomogeneity of the field $\mbox{\boldmath$B$}$\cite{PR65} or from a position-dependent $g$-factor.\cite{Salis01,JY01,JWSMS02,CBG05} Recently Kato {\it et al.}\cite{KMDGLA03} achieved operating electron spins in parabolic Al$_x$Ga$_{1-x}$As QWs through the $\hat g$-tensor modulation technique based on the difference in the spatial dependences of the various components of ${\hat g}={\hat g}(z)$. Graded parabolic QWs were originally designed for producing a high mobility electron gas \cite{H87,GradParab} by applying the modulation doping technique.\cite{ModLeg} The early work on their dynamical responses was mostly concerned with the effect of electron-electron interaction on the electron orbital dynamics and the related transition frequencies,\cite{BJH89} and also with generalizing the Kohn theorem.\cite{K61} The recent success in electrical spin manipulation\cite{KMDGLA03} shifted the interest to the spin flip transitions in such systems and the effect of various mechanisms of SO interaction.

Thd existence of several mechanisms of SO coupling makes it important to develop reliable experimental techniques for identifying them. Also, the relative efficiency of different SO coupling mechanisms strongly depends on the choice of the semiconductor materials and the shape of a QW or a heterojunction. The efficiency of $\hat g$-tensor modulation technique\cite{KMDGLA03} is based on the anomalously small $g$-factor value in Al$_x$Ga$_{1-x}$As, $|g|\alt 0.1$, and is therefore specific for GaAs based devices. Developing similar techniques for narrow-gap A$_3$B$_5$ semiconductors with typically large $g$ values, $|g|\agt 10$, needs different approaches, and we show that orbital mechanisms of SO coupling can be rather efficient for them. Moreover, we prove that the efficiency of different mechanisms depends strongly on the polarization of the ac electric field $\tilde{\mbox{\boldmath$E$}}$. 

To consider all these problems in a framework of an unified approach and to derive analytical expressions for the transition probabilities, we make two basic assumptions. First, we choose a parabolic shape for the QW. Second, we accept that SO coupling is not too strong, i.e., the SO coupling energy $E_{SO}$ is small compared with all different energies, including the confinement energy $\hbar\omega_0$, the cyclotron energy $\hbar\omega_c$, and the Zeeman energy $\hbar\omega_s$. Here $\omega_0$ is the characteristic frequency of the parabolic potential, $\omega_c=eB/mc$ is the  cyclotron frequency of electrons with the effective mass $m$, and $\omega_s=g\mu_BB/\hbar$ is the spin resonance frequency, with $\mu_B=e\hbar/2m_0c$ being the Bohr magneton. These assumptions allow one to account for ${\hat H}_{SO}$ only in the matrix elements of spin transitions and disregard the effect of ${\hat H}_{SO}$ on the position of energy levels.\cite{g-corr} In this approximation, one can take advantage of the exact solution\cite{Maan84,Mer87} for the quantization of electron levels in a magnetic field tilted with respect to the confinement plane. 

Meanwhile, applying the exact solution found by Maan\cite{Maan84} and Merlin\cite{Mer87} is not straightforward when it comes to calculating matrix elements of spin flip transitions. Indeed, their solution depends on an auxiliary angle $\gamma$ defined by the decoupling condition of two normal modes, hence, all matrix elements depend on $\gamma$. We have found an extensive set of remarkable identities relating $\gamma$, frequencies $\omega_{\xi,\eta}(\theta)$ of two eigenmodes, and the polar angle $\theta$. The symmetry of the problem underlying these identities is far from obvious, but the identities permitted us to eliminate $\gamma$ and find all final results in an explicit form. Some of these results, without the derivation, were published in our previous papers.\cite{RE03,APL04} We expect that the technique will facilitate developing the theory of different properties of parabolic QWs which is timely because of the recent progress in experimental work\cite{Salis01,KMDGLA03} and the competition of the different mechanisms of SO coupling that manifests itself in various phenomena.
 
In this paper we develop a general theory of EDSR in QWs which is caused by the standard Hamiltonians
of the ${\hat H}_{\rm orb}({\hat{\mbox{\boldmath$k$}}},\mbox{\boldmath$\sigma$})$ type for two basic geometries: with the ac electric field in the QW plane and perpendicular to this plane.  We solve exactly the problem of an electron confined in a parabolic QW being a subject to a tilted magnetic field $\mbox{\boldmath $B$}$ and find the EDSR intensity in the Dresselhaus and Rashba models {\it vs} the $\mbox{\boldmath $B$}$ direction. Our results show that electric dipole spin resonance is especially strong when it is excited by an in-plane electrical field. However, we show that it is also strong enough in the geometry when the time-dependent potential is applied to a gate. Our results demonstrate convincingly that an efficient electrical spin manipulation can be achieved through the orbital mechanisms of spin-orbit coupling. Our results also suggest that the angular dependence of EDSR intensity is an unique characteristic of various competing mechanisms of spin-orbit coupling  contributing to EDSR.

The paper is organized as follows. In Section II we develop a theoretical approach to an electron in a parabolic quantum well and derive operator expressions for the basic variables like the coordinates and kinetic momenta. While we apply these results for calculating the EDSR intensity, they are rather general and can be applied to different problems related to parabolic quantum wells. In Section III we calculate the EDSR intensity for the Dresselhaus and Rashba 2D spin orbit coupling Hamiltonians for the two basic geometries with an in-plane and perpendicular-to-plane electric field. We also investigate in detail the dependence of the EDSR intensity on the direction of the magnetic field and present figures that illustrate the basic mechanisms and characteristic features that should help assigning specific EDSR bands when experimental work begins. In Section IV we develop a theory of EDSR in a parabolic well with a 3D Dresselhaus SO interaction and unveil the specific features of EDSR that distinguish it from the results of the 2D model corresponding to a strong confinement limit. The discussion of obtained results and estimates of the EDSR intensity are given in Sect. V. Appendix A includes a number of identities that have been highly instrumental in deriving the results of Sections III and IV, and that we expect to be useful for future theoretical work on parabolic quantum wells. Appendix B supports the calculations of Sec. IV. 

\section{An electron in a parabolic well a subject to a tilted magnetic field}

Everywhere in what follows we suppose that SO corrections to the energy levels are small as compared with the separation between adjacent Landau levels and Zeeman sublevels. Therefore, we disregard the effect of SO coupling on the energy spectrum and begin with the Hamiltonian of the orbital motion in a parabolic quantum well
\be
{\hat H}_0={{\hbar^2}\over{2m}}(-i\mbox{\boldmath $\nabla$}+{e\over{\hbar c}}\mbox{\boldmath $A$})^2+{1\over2}m\omega_0^2z^2,
\label{eqn1}
\ee 
where $\mbox{\boldmath $A$}$ is the vector-potential of the magnetic field $\mbox{\boldmath $B$}$. Because the Hamiltonian is quadratic in the coordinates, the problem can be solved exactly, and the energy spectrum is well known.\cite{Maan84,Mer87} However, because for calculating matrix elements of EDSR we need explicit expressions for the operators of coordinates and momenta, somewhat lengthly calculations should be performed.

\subsection{Diagonalization of the Hamiltonian}

First, we perform a transformation $U$ from the original coordinate system $x,y,z$ related to the principal crystal axes to a new (``primed") reference system $x',y', z'$ with the $z'$ axis parallel to $\mbox{\boldmath $B$}$ and the $y'$ axis lying in the $x,y$ plane. The coordinates in both systems are related as:
\bea 
\left(\begin{array}{c}x\\y\\z
\end{array}\right)&=&\hat{U}\left(\begin{array}{c}x'\\y'\\z'\end{array}\right),~{\rm where}\nonumber\\ 
\hat{U}&=&\left( \begin{array}{ccc} \cos\theta\cos\phi&-\sin\phi&\sin\theta\cos\phi \\
 \cos\theta\sin\phi&\cos\phi&\sin\theta\sin\phi\\
 -\sin\theta&0& \cos\theta 
\end{array} \right).
\label{eqn2}
\eea
Similar equations are valid for momenta projections, $\hat{k}_{i}=\hat{U}_{ii'}\hat{k}_{i'}$, where $i=x,y,z$ and summation over repeated coordinate indices is implied. Because $y'$ does not appear in the potential energy $m\omega_0^2z^2/2$, it is convenient to choose the Landau gauge with $A_{x'}=0, A_{y'}=Bx', A_{z'}=0$. Then $y'$ is a cyclic variable,  the Landau momentum $k\equiv k_{y'}$ is a $c$-number, and
\bea 
{\hat H}_0&=&{{\hbar^2}\over{2m}}[-\partial^2_{x'}-\partial^2_{z'}+(k+x'/\lambda^2)^2]\nonumber\\
&+&{1\over2}m\omega_0^2(-\sin\theta~x'+\cos\theta~z')^2,
\label{eqn3}
\eea
where $\lambda=\sqrt{c\hbar/eB}$ is the magnetic length. To decouple the motion in two degrees of freedom in Eq.~(\ref{eqn3}), we perform a rotation
\bea
x'&=&\cos\gamma~\xi+\sin\gamma~ \eta,\nonumber\\
z'&=&-\sin\gamma~ \xi + \cos\gamma~ \eta.
\label{eqn4}
\eea
Then the cancelation condition for the mixed product $\xi\eta$ results in an equation on the auxiliary angle $\gamma$
\be
\sin{2\gamma}=(\omega_0/\omega_c)^2\sin[2(\theta+\gamma)],
\label{eqn5}
\ee
and the coefficients at $\xi^2$ and $\eta^2$, together with the kinetic energy, define the eigenfrequencies
\bea
\omega_\xi^2(\theta)&=&\omega_c^2\cos^2\gamma+\omega_0^2\sin^2(\theta+\gamma)\nonumber\\
 \omega_\eta^2(\theta)&=&\omega_c^2\sin^2\gamma+\omega_0^2\cos^2(\theta+\gamma)
\label{eqn6}
\eea
of two normal modes, $\xi$ and $\eta$. Equations (\ref{eqn5}) and (\ref{eqn6}) complete the diagonalization of the quadratic part of ${\hat H}_0$ and are in agreement with the results by Maan\cite{Maan84} and Merlin.\cite{Mer87}

Eliminating in Eq.~(\ref{eqn3}) the term that is linear in $x'$ can be achieved by shifting $\xi$ and $\eta$ by
\bea
\xi_0&=&{{-k\lambda^2\cos\gamma}\over{\cos^2\gamma+(\omega_0/\omega_c)^2\sin^2(\theta+\gamma)}}\nonumber\\
&=&-\lambda^2k~{{\cos(\theta+\gamma)}\over{\cos\theta}}~,\nonumber\\
\eta_0&=&{{-k\lambda^2\sin\gamma}\over{\sin^2\gamma+(\omega_0/\omega_c)^2\cos^2(\theta+\gamma)}}\nonumber\\
&=&-\lambda^2k~{{\sin(\theta+\gamma)}\over{\cos\theta}}.
\label{eqn7}
\eea
In these equations, first expressions for $\xi_0$ and $\eta_0$ come directly from calculations, while the simplified form of them can be found by using Eqs.~(\ref{eqA12}).

The free term in Eq.~(\ref{eqn3}) equals
\be
E_0(k)={{\hbar^2k^2}\over{2m}}-{m\over2}(\xi_0^2\omega_\xi^2+\eta_0^2\omega_\eta^2)=0.
\label{eqn8}
\ee
Vanishing of this term follows from Eqs.~(\ref{eqn5}) - (\ref{eqn7}), and independence of the energy from the Landau momentum $k$ is a consequence of the translational symmetry in the $(x,y)$ plane. The shifts $\xi_0,\eta_0$ satisfy the identities
\bea
\cos\gamma~\xi_0+\sin\gamma~\eta_0&=&-\lambda^2k,\nonumber\\
\sin(\theta+\gamma)~\xi_0-\cos(\theta+\gamma)~\eta_0&=&0\,\,.
\label{eqn9}
\eea

Equation~(\ref{eqn6}) allows one to find the spectrum of the system, i.e., the set of two frequencies, $\{\omega_\xi(\theta),\omega_\eta(\theta)\}$. However, because Eq.~(\ref{eqn6}) includes the auxiliary angle $\gamma$ that should be found from Eq.~(\ref{eqn5}), the shape of the curves  $\omega_\xi^2(\theta)$ and $\omega_\eta^2(\theta)$ is not obvious, depends on the ratio $\omega_c/\omega_0$, and identification of a specific eigenvalue as $\omega_\xi^2(\theta)$ or $\omega_\eta^2(\theta)$ is a matter of convention. In what follows, we identify $\omega_\xi(\theta)$ and $\omega_\eta(\theta)$ as the frequencies that at $\theta=0$ coincide with the cyclotron frequency $\omega_c$ and the confinement frequency $\omega_\eta$, respectively. Then, using Eqs.~(\ref{eqA13}) and (\ref{eqA15}), we derive explicit expressions for $\omega_\xi^2(\theta)$ and $\omega_\eta^2(\theta)$:
\bea
\omega_\xi^2(\theta)&=&{1\over2}\left(\omega_0^2+\omega_c^2-\Omega^2{\rm sign}\{\omega_0-\omega_c\}\right),\nonumber\\
\omega_\eta^2(\theta)&=&{1\over2}\left(\omega_0^2+\omega_c^2+\Omega^2{\rm sign}\{\omega_0-\omega_c\}\right),
\label{omsexpl}
\eea 
where
\be
\Omega^2(\theta)=\sqrt{\omega_0^4+\omega_c^4-2\omega_0^2\omega_c^2\cos2\theta}.
\label{Omega}
\ee

This spectrum is plotted in Fig.~\ref{fig:spectrum}a as a function of $\omega_c/\omega_0$ for three values of the polar angle $\theta$. For $\theta=0$, the spectrum consists of the cyclotron and confinement branches, $\xi$ and $\eta$, respectively. However, for $\theta\neq0$ the branches interchange at $\omega_c=\omega_0$. Left parts of the lower branches and right parts of the upper branches are described by the $\omega_\xi$ solution and show a cyclotron-like behavior. The opposite parts of the same branches are described by the $\omega_\eta$ solution and show a confinement-like behavior. Of course, all branches are continuous and smooth.

The same spectrum is plotted in Fig.~\ref{fig:spectrum}b as the function of $\theta$ for three values of $\omega_c/\omega_0$. All branches retain their identity, $\xi$ or $\eta$, in the whole interval $0\leq\theta\leq\pi/2$. All $\eta$ branches originate at the same frequency $\omega_\eta(\theta=0)=\omega_0$; they are upper branches for $\omega_c<\omega_0$ and lower branches for $\omega_c>\omega_0$. On the contrary, $\xi$ branches are lower branches for $\omega_c<\omega_0$ and upper branches for $\omega_c>\omega_0$; they originate at the frequencies $\omega_c$ that are $B$-dependent. For $\omega_c=\omega_0$, the spectrum is described by a simple equation $\omega_\pm(\theta)=\omega_0(1\pm\sin\theta)^{1/2}$. These curves are separatrices dividing $\xi$ and $\eta$ regions, hence, neither $\xi$ nor $\eta$ identity can be ascribed to them.

The Hamiltonian ${\hat H}_0$ of Eq.~(\ref{eqn3}), when written in the variables $\xi$ and $\eta$, reads
\be
{\hat H}_0=\sum_\zeta[-{{\hbar^2}\over{2m}}\partial_\zeta^2+{1\over2}m\omega_\zeta^2(\zeta-\zeta_0)^2],
\label{eqn10}
\ee 
where the summation is performed over $\zeta=\xi,\eta$. After introducing step operators
\bea
\zeta-\zeta_0&=&\sqrt{\hbar/2m\omega_\zeta}(a_\zeta^++a_\zeta)~,\nonumber\\
{\hat k}_\zeta&=&-i\partial_\zeta=i\sqrt{m\omega_\zeta/2\hbar}(a_\zeta^+-a_\zeta)~,
\label{eqn11}
\eea
the Hamiltonian takes the standard oscillator form
\be
{\hat H}_0={1\over2}\sum_\zeta\hbar\omega_\zeta(a_\zeta^+a_\zeta + a_\zeta a_\zeta^+).
\label{eqn12}
\ee

\subsection{Operator representation for coordinates and momenta}
\label{operators}

Similarly to Eq.~(\ref{eqn2}), one can express the components ${\hat k}_j=-i\nabla_j+(e/\hbar c)A_j$ of the kinetic momentum in the crystal frame through its components in the primed frame. In the latter frame, the vector-potential 
\mbox{\boldmath$A$} contributes only to the component ${\hat k}_{y'}$ that equals
\be
{\hat k}_{y'}=k+x'/\lambda^2=[\cos\gamma~(\xi-\xi_0)+\sin\gamma~(\eta-\eta_0)]/\lambda^2.
\label{eqn13}
\ee
When deriving Eq.~(\ref{eqn13}), the upper identity of Eq.~(\ref{eqn9}) has been used. Using Eq.~(\ref{eqn4}), one can express ${\hat k}_{x'}$ and ${\hat k}_{z'}$ through the derivatives $\partial_\zeta$ and rewrite the components of the momenta in the crystal frame in terms of $a_\zeta,a_\zeta^+$ by applying Eq.~(\ref{eqn11}) 
\bea
\hat{k}_x&=&\sum_\zeta (X_\zeta a_\zeta+X_\zeta^*a^+_\zeta),\nonumber\\
 \hat{k}_y&=&\sum_\zeta (Y_\zeta a_\zeta+Y_\zeta^*a^+_\zeta),\nonumber\\
\hat{k}_z&=&\sum_\zeta (Z_\zeta a_\zeta+Z_\zeta^*a^+_\zeta).
\label{eqn14}
\eea
Here
\bea
X_\xi&=&\sqrt{m\omega_\xi\over 2\hbar}\left[-i\cos(\theta+\gamma)\cos\phi- {\omega_c\over \omega_\xi}\cos\gamma\sin\phi\right]~,\nonumber\\
X_\eta&=&\sqrt{m\omega_\eta\over 2\hbar}\left[-i\sin(\theta+\gamma)\cos\phi- {\omega_c\over\omega_\eta}\sin\gamma\sin\phi\right]~,\nonumber\\
Y_\xi&=&-\sqrt{m\omega_\xi\over 2\hbar}\left[i\cos(\theta+\gamma)\sin\phi-{\omega_c\over\omega_\xi}\cos\gamma\cos\phi\right]~,\nonumber\\
Y_\eta&=&-\sqrt{m\omega_\eta\over 2\hbar}\left[i\sin(\theta+\gamma)\sin\phi-{\omega_c\over\omega_\eta}\sin\gamma\cos\phi\right]~,\nonumber\\
Z_\xi&=&i\sqrt{m\omega_\xi\over 2\hbar}\sin(\theta+\gamma)~,\nonumber\\
Z_\eta&=&-i\sqrt{m\omega_\eta\over 2\hbar}\cos(\theta+\gamma)~.
\label{eqn15}
\eea

These coefficients depend only on the angles $(\theta, \phi)$ and the eigenfrequencies $\omega_\zeta$ and do not depend on $k$, $\xi_0$, and $\eta_0$. Therefore, despite the fact that Eqs.~(\ref{eqn14}) and (\ref{eqn15}) have been derived in the Landau representation, the final results are gauge invariant and may be conveniently employed for finding operators  $({\hat x},{\hat y},{\hat z})$ of the coordinates $(x,y,z)$.

It is seen from Eq.~(\ref{eqn1}) that the operator $\hat z$ can be found in terms of $(a_\zeta,a_\zeta^+)$ from the difference $H_0-\hbar^2{\hat{\mbox{\boldmath$k$}}}^2/2m$. Employing Eqs.~(\ref{eqn12}), (\ref{eqn14}), and (\ref{eqn15}) and some of the identities of Appendix A, we arrive after somewhat lengthly algebra at
\be
\hat{z}=\sum_\zeta(i\hbar/m\omega_\zeta)Z_\zeta(a_\zeta+a_\zeta^+).
\label{eqn16}
\ee
The sign of $\hat z$ has been found from the commutation relation $[{\hat k}_z,{\hat z}]=-i$, and one can easily check that ${\hat{\dot z}}=(i/\hbar)[{\hat H}_0,{\hat z}]=\hbar{\hat k}_z/m$.

For finding $\hat x$ and $\hat y$, we generalize the Johnson and Lippman \cite{JL49} procedure for the quantization of electron motion in a strong magnetic field for an electron confined in a parabolic quantum well. Using the equations of motion for the operators of momenta and coordinates
\be
{\hat{\dot k}}_x=\omega_c(b_y{\hat k}_z-b_z{\hat k}_y),\,{\hat{\dot k}}_y=\omega_c(b_z{\hat k}_x-b_x{\hat k}_z),
\label{eqn17}
\ee
\be
{\dot x}=\hbar {\hat k}_x/m,\, {\dot y}=\hbar {\hat k}_y/m,\, {\dot z}=\hbar {\hat k}_z/m,
\label{eqn18}
\ee
where $\mbox{\boldmath$b$}=\mbox{\boldmath$B$}/B$, and eliminating the components of the momentum $\hat{\mbox{\boldmath$k$}}$, we arrive at the equations
\be
{\hat{\dot x}}=(\lambda^2{\hat{\dot k}}_y+b_x{\hat{\dot z}})/b_z,\,\,
{\hat{\dot y}}=(-\lambda^2{\hat{\dot k}}_x+b_y{\hat{\dot z}})/b_z.
\label{eqn19}
\ee
Equations (\ref{eqn19}) suggest existence of two integrals of motion
\be
{\hat x}_0={\hat x}-(b_x {\hat z}+\lambda^2{\hat k}_y)/b_z,\,\,
{\hat y}_0={\hat y}-(b_y {\hat z}-\lambda^2{\hat k}_x)/b_z,
\label{eqn20}
\ee
that generalize the well known guiding center coordinates. They obey the same commutation relation 
\be
[{\hat x}_0,{\hat y}_0]=i\lambda^2/\cos\theta
\label{commrel}
\ee 
and commute with operators $a_\zeta$. Remarkably, in a parabolic well the operators ${\hat x}_0$ and ${\hat y}_0$ involve the vertical coordinate $\hat z$. Nevertheless, Eq.~(\ref{commrel}) ensures that the degeneracy of states is completely controlled by $B_z$, the component of $\mbox{\boldmath$B$}$ perpendicular  to the QW plane, and is described by the standard Landau formula $n_L(\theta)=\cos\theta/2\pi\lambda^2$.

Because the operators ${\hat k}_x$, ${\hat k}_y$, and $\hat z$ are already known, Eq.~(\ref{eqn20}) allows one to find $\hat x$ and $\hat y$. To derive explicit expressions for these operators, it is convenient to use identities
\bea
Y_\zeta+(i\omega_c/\omega_\zeta)\sin\theta\cos\phi Z_\zeta&=&(i\omega_c/\omega_\zeta)\cos\theta X_\zeta\,,\nonumber\\
X_\zeta+(i\omega_c/\omega_\zeta)\sin\theta\sin\phi Z_\zeta&=&(i\omega_c/\omega_\zeta)\cos\theta Y_\zeta
\label{eqn21}
\eea
that can be checked using Eqs.~(\ref{eqn6}) and (\ref{eqn15}). Finally, the operators of in-plane coordinates are
\bea
{\hat x}&=&\sum_\zeta(i\hbar/m\omega_\zeta)(X_\zeta a_\zeta-X_\zeta^*a_\zeta^+)+{\hat x}_0,\nonumber\\
{\hat y}&=&\sum_\zeta(i\hbar/m\omega_\zeta)(Y_\zeta a_\zeta-Y_\zeta^*a_\zeta^+)+{\hat y}_0.
\label{eqn21}
\eea
Because $Z_\zeta^*=-Z_\zeta$, Eqs.~(\ref{eqn21}) are similar to Eq.~(\ref{eqn16}) and differ from it only by presence of the guiding center operators.

Therefore, six operators of coordinates and momenta, $x_j$ and ${\hat k}_j$, $j=x,y,z$, are expressed in terms of four Bose operators $(a_\zeta,a_\zeta^+)$ and two generalized guiding center coordinates $({\hat x}_0,{\hat y}_0)$. The latter commute with all operators $(a_\zeta,a_\zeta^+)$; hence, their presence in Eq.~(\ref{eqn21}) does not influence electron dynamics in homogeneous external fields that are coupled to an electron only through the operators $(a_\zeta,a_\zeta^+)$. Operators ${\hat x}_j$ and ${\hat k}_j$, defined by Eqs.~(\ref{eqn14}), (\ref{eqn16}), and (\ref{eqn21}), obey the standard commutation relations
\be
[{\hat k}_j,{\hat x}_\ell]=-i\delta_{j\ell},\,\,[{\hat k}_j,{\hat k}_\ell]=-i\lambda^{-2}\epsilon_{j\ell m}B_m/B,
\label{eqn22}
\ee
where $\epsilon_{j\ell m}$ is the Levi-Civita tensor.

\section{Intensity of EDSR}
\label{EDSR}
 
The EDSR occurs in QWs due spin-orbit Hamiltonians $\hat{H}_{SO}$ described by Eqs.~(\ref{eq11}) - (\ref{eqR}) which  mix the electron spin projections on the magnetic field direction, hence, each electron state acquires admixture of the opposite spin projection that is small when ${\hat H}_{SO}$ is weak. As a result, a time dependent electric field $\tilde{E}(t)$ causes EDSR, i.e., electrically-induced spin-flip transitions. Matrix elements of EDSR are comprised of two contributions, of which one comes from the perturbation of the wave functions of stationary states and the second from the direct coupling of the electron spin to the field $\tilde{E}(t)$. A convenient way for calculating EDSR matrix elements is based on employing a canonical transformation $\exp({\hat T})$ eliminating ${\hat H}_{\rm so}$ in the first order of perturbation theory.\cite{RS61,RS91} After the transformation, the time-independent part of $\hat{H}$ conserves the electron spin projection on the magnetic field direction, and spin-flip transitions are induced only by the spin-orbit contribution $e(\hat{\mbox{\boldmath $v$}}_{\rm so}\cdot\mbox{\boldmath $\tilde{A}$})/c$ to the time-dependent part of $\hat{H}$, where $\mbox{\boldmath $\tilde{A}$}(t)$ is a time-dependent vector potential, and $\hat{\mbox{\boldmath $v$}}_{\rm so}$ is the spin-orbit part of the velocity operator. 

Because of the equation of motion $\hat{\mbox{\boldmath$v$}}=(i/\hbar)[{\hat H},\hat{\mbox{\boldmath$r$}}]$, the matrix elements of the operators $\hat{\mbox{\boldmath$v$}}$ and $\hat{\mbox{\boldmath$r$}}$ for spin-flip transitions are related as $\langle\hat{\mbox{\boldmath$v$}}\rangle=i\omega_s\langle\hat{\mbox{\boldmath$r$}}\rangle$, with $\omega_s$ being the spin-flip transition frequency. Neither of the two competing contributions to $\hat{\mbox{\boldmath$v$}}$ includes the factor $\omega_s$, hence, it indicates the existence of massive cancellations that tremendously complicate calculations based on the operator $\hat{\mbox{\boldmath$v$}}$. It is much more convenient to write the time dependent part of the Hamiltonian as $\hat{H}_{\rm int}(t)=e(\hat{\mbox{\boldmath$r$}}\cdot{\tilde{\mbox{\boldmath$E$}}}(t))$. In the original representation, the coordinate operator $\hat{\mbox{\boldmath$r$}}$ is diagonal in spin indices, and $\hat{H}_{\rm int}(t)$ produces spin-flip transitions due to the level mixing. However, after the $\hat T$-transformation the operator $\hat{\mbox{\boldmath$r$}}$ acquires a SO part $\hat{\mbox{\boldmath$r$}}_{\rm so}=[\hat{T},\hat{\mbox{\boldmath$r$}}]$ that drives spin-flip transitions. 
 
The total Hamiltonian of an electron confined in a parabolic QW is $\hat{H}=\hat{H}_0+\hat{H}_Z+\hat{H}_{\rm so}+\hat{H}_{\rm int}(t)$, with $\hat{H}_Z=\frac{1}{2}g\mu_B(\mbox{\boldmath $\sigma$}\cdot\mbox{\boldmath $B$})$. The energy levels of the Hamiltonian $\hat{H}_0+\hat{H}_Z$ are
\be
E_\sigma(n_\xi, n_\eta)=\sum_\zeta\hbar\omega_\zeta(\theta)(n_\zeta+1/2)+\hbar\omega_s\sigma/2\,,
\label{spectrum}
\ee
where $n_{\xi,\eta}\geq 0$ and the spin index $\sigma=\pm 1$. The spin-flip frequency $\omega_s$ should be taken algebraically; $\omega_s<0$ for electrons in negative $g$-factor semiconductors. The time independent spin-orbit interaction $\hat{H}_{\rm so}$ will be considered as a perturbation.

The term ${\hat H}_{\rm so}= \hat{\cal H}_D,~{\hat H}_D,~{\rm or}~{\hat H}_R $ in the time-independent part of the Hamiltonian $\hat{H}$ leads to the mixing of spin sublevels. As we have stated  above, it can be  eliminated  in the first order of the perturbation theory by a canonical transformation $\exp({\hat T})$.  The operator ${\hat T}$ is nondiagonal in the orbital quantum numbers $(n_\xi, n_\eta)$, and its matrix elements are 
\be
\langle n_\xi^\prime, n_\eta^\prime, \sigma^\prime\vert {\hat T}\vert n_\xi, n_\eta, \sigma\rangle={{\langle n_\xi^\prime, n_\eta^\prime, \sigma^\prime\vert {\hat H}_{\rm so}\vert n_\xi, n_\eta, \sigma\rangle}\over{E_{\sigma^\prime}(n_\xi^\prime, n_\eta^\prime)-E_\sigma(n_\xi, n_\eta)}}\,.
\label{Tmatrix}
\ee
In terms of $\hat T$, the matrix elements of spin-flip transitions diagonal in $(n_\xi,n_\eta)$ are
\bea
&&\langle n_\xi,n_\eta,\uparrow\vert(\tilde{\mbox{\boldmath$E$}}\cdot\hat{\mbox{\boldmath$r$}}_{\rm so})\vert n_\xi,n_\eta,\downarrow\rangle\nonumber\\
&=&\sum_{n_\xi',n_\eta'}\{\langle n_\xi,n_\eta,\uparrow|\hat{T}|n_\xi',n_\eta',\downarrow\rangle\langle n_\xi',n_\eta'|(\tilde{\mbox{\boldmath$E$}}\cdot\hat{\mbox{\boldmath$r$}})|n_\xi,n_\eta\rangle\nonumber\\
&-&\langle n_\xi,n_\eta|(\tilde{\mbox{\boldmath$E$}}\cdot\hat{\mbox{\boldmath$r$}})|n_\xi',n_\eta'\rangle\langle n_\xi',n_\eta',\uparrow|\hat{T}|n_\xi,n_\eta,\downarrow\rangle\}~,
\label{transition}
\eea
with $n_\xi'=n_\xi\pm1$ and $n_\eta'=n_\eta\pm1$. The sum  in Eq.~(\ref{transition}) is restricted because the operator $\hat{\mbox{\boldmath$r$}}$ is linear in $(a_\zeta,a_\zeta^+)$, cf. Eqs.~(\ref{eqn16}) and (\ref{eqn21}). This restriction significantly simplifies matrix elements $\langle n_\xi',n_\eta',\uparrow|\hat{T}|n_\xi,n_\eta\downarrow\rangle$ since their denominators do not depend on $(n_\xi,n_\eta)$ and are equal to $\hbar(\pm\omega_\zeta+\omega_s)$. 

When the operator $\hat{H}_{\rm SO}$ is linear in momenta and consequently in the operators $(a_\zeta,a_\zeta^+)$, Eq.~(\ref{transition}) reduces to 
\bea
&&\langle n_\xi,n_\eta,\uparrow\vert(\tilde{\mbox{\boldmath$E$}}\cdot\hat{\mbox{\boldmath$r$}}_{\rm so})\vert n_\xi,n_\eta,\downarrow\rangle\nonumber\\
&=&-\sum_{\zeta=\xi,\eta}\left[{(\tilde{\mbox{\boldmath$E$}}\cdot{\mbox{\boldmath$l$}^*}_\zeta)\langle 0_\zeta,\uparrow|{\hat H}_{\rm so}
|1_\zeta,\downarrow\rangle\over \hbar(\omega_\zeta-\omega_s)}\right.\nonumber\\
&+&\left.{(\tilde{\mbox{\boldmath$E$}}\cdot{\mbox{\boldmath$l$}}_\zeta)\langle 1_\zeta,\uparrow|{\hat H}_{\rm so}
|0_\zeta,\downarrow\rangle\over \hbar(\omega_\zeta+\omega_s)}\right]
\label{melement1}
\eea
with
\be 
l_{x\zeta}={i\hbar\over m\omega_\zeta}X_\zeta,~l_{y\zeta}={i\hbar\over m\omega_\zeta}Y_\zeta,~l_{z\zeta}={i\hbar\over m\omega_\zeta}Z_\zeta.
\label{ls}
\ee
Remarkably, this matrix element does not depend on $(n_\xi,n_\eta)$. For calculating the matrix element $\langle \uparrow|{\hat H}_{\rm so}|\downarrow\rangle$ in Eq.~(\ref{melement1}), we transform Pauli matrices from the crystal frame $(x,y,z)$ to the primed frame $(x',y',z')$ as  
$\hat{\sigma}_{i}=\hat{U}_{ii'}\hat{\sigma}_{i'}$,
similarly to Eq.~(\ref{eqn2}).  Substituting $\hat{\sigma}_{x,y}$ into the 2D Dresselhaus and Rashba Hamiltonians of Eqs.~(\ref{eqDr}) and (\ref{eqR}), and using $\langle \uparrow|{\hat \sigma}_{x'}|\downarrow\rangle=1$, $\langle \uparrow|{\hat \sigma}_{y'}|\downarrow\rangle=-i$, and $\langle \uparrow|{\hat \sigma}_{z'}|\downarrow\rangle=0$, we arrive at
\be
\langle \uparrow|{\hat H}_{\rm so}|\downarrow\rangle=\sum_{\zeta}(H^-_\zeta\hat{a}_\zeta+H^+_\zeta\hat{a}_\zeta^+)~, 
\label{eqn35}
\ee
with the matrix elements $H^\pm_\zeta$ that should be found from Eqs.~(\ref{eqn14}) and (\ref{eqn15}).
Applying Eq.~(\ref{eqA12}), we find for a Rashba Hamiltonian
\be
H^{\mp}_{R,\zeta}=\alpha_R\sqrt{m\omega_\zeta/ 2\hbar}~\left({\omega_\zeta/\omega_c}\pm 1\right)f_\zeta~,
\label{HR}
\ee
with $f_\xi=\cos(\theta+\gamma)$ and $f_\eta= \sin(\theta+\gamma)$, and for a 2D Dresselhaus Hamiltonian 
\bea
H^{\mp}_{2D,\zeta}&=&\alpha_D\sqrt{m\omega_\zeta\over 2\hbar}\left[i\cos2\phi\left({\omega_\zeta\over \omega_c\cos\theta}\mp\cos\theta\right)\right.\nonumber\\
&-&\left.\sin2\phi\left({\omega_\zeta\over \omega_c} \mp 1\right)\right]f_\zeta~. 
\label{HD}
\eea
Substituting Eq.~(\ref{eqn35}) into Eq.~(\ref{melement1}), we find
\bea
&&\langle n_\xi,n_\eta,\uparrow\vert(\tilde{\mbox{\boldmath$E$}}\cdot\hat{\mbox{\boldmath$r$}}_{\rm so})\vert n_\xi,n_\eta,\downarrow\rangle=-{1\over \hbar}\sum_{\zeta}{1\over \omega_\zeta^2-\omega_s^2}\nonumber\\
&\times&\{\omega_\zeta[H^-_\zeta({\mbox{\boldmath$l$}}_\zeta\cdot \tilde{\mbox{\boldmath$E$}})^*+H^+_\zeta({\mbox{\boldmath$l$}}_\zeta\cdot \tilde{\mbox{\boldmath$E$}})] \nonumber\\
&+&\omega_s[H^-_\zeta({\mbox{\boldmath$l$}}_\zeta\cdot \tilde{\mbox{\boldmath$E$}})^*-H^+_\zeta({\mbox{\boldmath$l$}}_\zeta\cdot \tilde{\mbox{\boldmath$E$}})]\}
\label{melement2}
\eea

\subsection{Electric field perpendicular to the QW plane}
\label{sec:perp}

Let us start with a time-dependent electric field perpendicular to the QW plane, $\mbox{\boldmath $\tilde{E}$}(t)\parallel{\bf\hat{z}}$.  In this geometry $l_{z,\zeta}= l_{z,\zeta}^*$, and the matrix element of Eq.~(\ref{melement2}), $L^z=\langle n_\xi,n_\eta,\uparrow\vert\hat{z}_{\rm so}\vert n_\xi,n_\eta,\downarrow\rangle$, equals
\be
L^z=-{1\over \hbar}\sum_\zeta l_{z\zeta}\left[ \omega_\zeta{H^-_{\zeta}+H^+_{\zeta} \over \omega_\zeta^2-\omega_s^2} +\omega_s{H^-_{\zeta}-H^+_{\zeta}\over \omega_\zeta^2-\omega_s^2}\right].
\label{Lz}
\ee
For a Rashba SO coupling, substituting Eq.~(\ref{HR}) into Eq.~(\ref{Lz}) results in 
\be
L^z_R={\alpha_R\over2\hbar}\left(1+{\omega_s\over\omega_c}\right)
{\omega_s(\omega_\eta^2-\omega_\xi^2)\over {\cal D}(\theta)}\sin2(\theta+\gamma)~,
\ee
with
\bea
{\cal D}(\theta)&\equiv&(\omega^2_\xi-\omega_s^2)(\omega^2_\eta-\omega_s^2)\nonumber\\
&=&\omega_0^2\omega_c^2\cos^2\theta-\omega_s^2(\omega_0^2+\omega_c^2-\omega_s^2);
\label{denom}
\eea
here, Eq.~(\ref{eqA9}) has been applied. By using Eq.~(\ref{eqA14}), one arrives at the final result\cite{RE03,error}
\be
L^z_R=-{\alpha_R\omega_s\over 2\hbar}{(\omega_c+\omega_s)\omega_c
\over {\cal D}(\theta)}\sin2\theta~.
\label{R2D}
\ee
For Dresselhaus SO coupling, a similar procedure leads to the matrix element of Ref.~[\onlinecite{RE03}]
\bea
L^z_D&=&-{\alpha_D\omega_s \over \hbar}{\omega_c\sin\theta\over {\cal D}(\theta)}\left[ \sin2\phi\cos\theta\left(\omega_c-\omega_s\right)\right.\nonumber\\
&-&i\cos2\phi\left.\left(\omega_c\cos^2\theta-\omega_s\right)\right]~.
\label{D2D}
\eea

\subsection{Electric field in the quantum well plane}
\label{sec:inplane}

 For an in-plane electric field, we calculte the scalar product $({\mbox{\boldmath$l$}}_\zeta\cdot \tilde{\mbox{\boldmath$E$}})$. If the field $\tilde{\mbox{\boldmath$E$}}(t)$ is polarized at the angle $\psi$ to the crystal $x$-axis, $\tilde{\mbox{\boldmath$E$}}(t)=\tilde{E}(t)(\cos\psi, \sin\psi, 0)$, then $({\mbox{\boldmath$l$}}_\zeta\cdot \tilde{\mbox{\boldmath$E$}})=l_\zeta^{\psi}\tilde{E}(t)$ with
\bea
l_\zeta^{\psi}&=&l_{x\zeta}\cos\psi+l_{y\zeta}\sin\psi=\sqrt{\hbar\over 2m\omega_\zeta}f_\zeta\nonumber\\
&\times&\left[\cos(\phi-\psi)-i{\omega_\zeta\over \omega_c\cos\theta}\sin(\phi-\psi)\right]~.
\label{ED}
\eea
When deriving Eq.~(\ref{ED}), the identity of Eq.~(\ref{eqA6}) has been used. Equation (\ref{ED}) allows one to rewrite the matrix element $L^\psi=\langle n_\xi,n_\eta,\uparrow\vert\hat{x}_{\rm so}\cos\psi+\hat{y}_{\rm so}\sin\psi\vert n_\xi,n_\eta,\downarrow\rangle$  of Eq.~(\ref{melement1}) as 
\bea
L^\psi&=&-{1\over \hbar}\sum_{\zeta} {\rm Re}\{l_{\zeta}^\psi\}\left[ \omega_\zeta{H^-_{\zeta}+H^{+}_{\zeta} \over \omega_\zeta^2-\omega_s^2} +\omega_s{H^-_{\zeta}-H^{+}_{\zeta}\over \omega_\zeta^2-\omega_s^2}\right]\nonumber\\
&+&{i\over \hbar}\sum_\zeta {\rm Im}\{l_{\zeta}^\psi\}\left[ \omega_s{H^-_{\zeta}+H^{+}_{\zeta} \over \omega_\zeta^2-\omega_s^2} +\omega_\zeta{H^-_{\zeta}-H^{+}_{\zeta}\over \omega_\zeta^2-\omega_s^2}\right]\,.\nonumber\\
\label{Lpsi}
\eea
For Rashba SO coupling, using Eq.~(\ref{HR}) results in
\bea
L^\psi_R&=&-{\alpha_R\cos(\phi-\psi)\over \hbar}\sum_\zeta\left[ {1\over \omega_c}{f_\zeta^2\omega_\zeta^2\over \omega_\zeta^2-\omega_s^2}+ \omega_s {f_\zeta^2\over \omega_\zeta^2-\omega_s^2}\right]\nonumber\\
&-&i{\alpha_R\sin(\phi-\psi)\over \cos\theta\hbar\omega_c}\left(1+{\omega_s\over \omega_c}\right)\sum_\zeta {f_\zeta^2\omega_\zeta^2\over \omega_\zeta^2-\omega_s^2}~.
\label{LpsiR}
\eea
Applying identities of Eq.~(\ref{eqA3}), one finds an explicit expression for the dependence of the matrix element on the magnetic field direction
\bea
L^\psi_R&=&-{\alpha_R\over \hbar}\cos(\phi-\psi)\nonumber\\
 &\times&{\omega_c\cos^2\theta(\omega_0^2-\omega_s^2)+\omega_s(\omega_0^2+\omega_c^2\sin^2\theta-\omega_s^2)\over{\cal D}(\theta)}\nonumber\\
&-&i{\alpha_R \over \hbar}\cos\theta\sin(\phi-\psi)~{ (\omega_c+\omega_s)(\omega_0^2-\omega_s^2)\over{\cal D}(\theta)}~.
\label{LpsiR1}
\eea
 In the strong 2D limit, when $\omega_0\gg \omega_c$, this expression simplifies significantly, and we arrive at the result of Ref.~[\onlinecite{APL04}]
\bea
L^\psi_R&=&-{\alpha_R\over \hbar(\omega^{*2}_c-\omega_s^2)}[\cos(\phi-\psi)(\omega^*_c\cos\theta+\omega_s)\nonumber\\
&+&i\sin(\phi-\psi)(\omega^*_c+\omega_s\cos\theta)]~.
\label{APLR}
\eea
Here $\omega_c^*=\omega_c\cos\theta $ is the cyclotron frequency in a tilted magnetic field; from now on $\theta$ is defined as $0\leq\theta\leq\pi/2$.

For a 2D Desselhaus SO coupling, substituting Eq.~(\ref{HD}) into Eq.~(\ref{Lpsi}) results in
\bea
&&L^\psi_D/\alpha_D=\nonumber\\
&-&{\cos(\phi-\psi)\over \hbar}\sum_\zeta\left[ {f_\zeta^2\omega_\zeta^2\over \omega_\zeta^2-\omega_s^2}{i\cos2\phi -\sin2\phi\cos\theta\over \omega_c\cos\theta}\right.\nonumber\\
&+& \left. {\omega_s f_\zeta^2\over \omega_\zeta^2-\omega_s^2}(\sin2\phi -i\cos2\phi\cos\theta)\right]\nonumber\\
&-&{i\sin(\phi-\psi)\over \cos\theta\hbar\omega_c}\sum_\zeta {f_\zeta^2\omega_\zeta^2\over \omega_\zeta^2-\omega_s^2}\left[{\omega_s(i\cos2\phi -\sin2\phi\cos\theta)\over \omega_c\cos\theta}\right.\nonumber\\
&+&\left.\sin2\phi -i\cos2\phi\cos\theta\right]~.
\label{LpsiD}
\eea
By using identities of Eq.~(\ref{eqA3}), an explicit expression for this matrix element can be found 
\bea
&&L^\psi_D/\alpha_D=-{\cos(\phi-\psi)\over \hbar}\left[ { \omega_c\cos\theta(\omega_0^2-\omega_s^2)\over{\cal D}(\theta)}\right.\nonumber\\
&\times&(i\cos2\phi -\sin2\phi\cos\theta)\nonumber\\
 &+&\omega_s\left.{(\omega_0^2+\omega_c^2\sin^2\theta-\omega_s^2)\over{\cal D}(\theta)}(\sin2\phi -i\cos2\phi\cos\theta)\right]\nonumber\\
&-&{i\sin(\phi-\psi)\over \hbar}{ (\omega_0^2-\omega_s^2)\over{\cal D}(\theta)}\left[\omega_s(i\cos2\phi -\sin2\phi\cos\theta)\right.\nonumber\\
&+&\left.\omega_c\cos\theta(\sin2\phi -i\cos2\phi\cos\theta)\right]~.
\label{LpsiD1}
\eea

 In the strong 2D limit, $\omega_0\gg \omega_c$, through a number of transformations this expression can be brought to the final form\cite{APL04}
\bea
L^\psi_D&=&{\alpha_D\over \hbar(\omega^{*2}_c-\omega_s^2)}[\sin(\phi+\psi)(\omega_c^*\cos\theta-\omega_s)\nonumber\\
&-&i\cos(\phi+\psi)(\omega_c^*-\omega_s\cos\theta)]\,.
\label{APLD}
\eea

In conclusion, we note that in this and the previous subsection we considered the Hamiltonians ${\hat H}_D$ and ${\hat H_R}$ on a similar footing. However, there is a considerable difference in their ranges of applicability. The Hamiltonian ${\hat H}_D$ originates from the 3D Hamiltonian ${\hat{\cal H}}_D$ and is valid only in the strong confinement limit when $d\ll k_F^{-1},\lambda$, where $k_F$ is the Fermi momentum.  When the confinement length $d$ becomes comparable to $k_F^{-1}$ or the magnetic length $\lambda$, additional terms like $k_xk_y^2$ should be included in the 2D Hamiltonian. From this standpoint, only Eq.~(\ref{APLD}) has a real physical meaning while Eq.~(\ref{LpsiD1}) should be only considered as an auxiliary one. Having those limitations in mind, Eq.~(\ref{LpsiD1}) will be applied in Sec.~\ref{sec:Azimuth} only to electrons confined at the ground level, $n_\xi,n_\eta=0$. Eq.~(\ref{D2D}) is a subject to similar restrictions, and EDSR with $n_\xi,n_\eta\neq0$ electrons in a field ${\tilde{\mbox{\boldmath$E$}}}\parallel{\hat{\bf z}}$ will be considered in Sec.~\ref{sec:3D} in the framework of a more general theory. 

The applicability range of the Hamiltonian ${\hat H_R}$ is much wider because it is applicable not only to zinc blende crystals under the conditions of 2D confinement but is inherent in the wurtzite modification of A$_3$B$_5$ compounds as a bulk property.\cite{R60,RS91} Currently, active experimental work on SO properties of microstructures including the wurtzite modifications of InAs,\cite{InAs} InN,\cite{InN}  GaN,\cite{GaN1,GaN2,GaN3} etc. is under way, and large SO splittings up to 9 meV have been reported.\cite{GaN2} Following the initial calculations of the band structure of wurtzite type compounds,\cite{Cardona} more general models have been developed recently,\cite{L03,Voon} and it looks like  the band folding in the hexagonal direction (resulting from the different size of the elementary cells in the zinc blende and wurtzite lattices) plays a role in developing large spin splittings.\cite{Lo05} Therefore, the Hamiltonian ${\hat H_R}$ will be applied for arbitrary $n_\xi,n_\eta\neq0$.

\subsection{Polarization dependence of EDSR intensity: General properties}
\label{PolarDepen}

In this section, we discuss at a qualitative level the basic properties of EDSR following from the equations of Sections \ref{sec:perp} and \ref{sec:inplane}. 

In all cases, the intensity of EDSR shows a pole when one of the eigenfrequencies coincides with $|\omega_s|$, $\omega_\zeta(\theta_{\rm pole})=|\omega_s|$. It is seen from Fig.~\ref{fig:spectrum}b that such a pole always exists because one of the frequencies $\omega_\zeta(\theta)$ vanishes for $\theta=\pi/2$.\cite{semimag} For $\omega_0>\omega_c$ it happens to the $\omega_\xi$ mode while for $\omega_0<\omega_c$ to the $\omega_\eta$ mode. The magnitude of $\theta_{\rm pole}$ depends on the relative magnitudes of $\omega_0$, $\omega_c$, and $\omega_s$. For $\omega_0\gg\omega_c$, $\theta_{\rm pole}$ depends mostly on the ratio $\omega_s/\omega_c$. It is very close to $\pi/2$ for GaAs because of its anomalously small $g$-factor, and equals $\theta_{\rm pole}\approx 0.89\times\pi/2$ for InAs and $\theta_{\rm pole}\approx 0.78\times\pi/2$ for InSb. Therefore, for the materials like InAs and InSb the maximum can still be achieved  in the range of $\theta$ and $B$ values where the transformation of Eq.~(\ref{Tmatrix}) remains justified. Additional dependence of the matrix elements on $\theta$ comes from the numerators of Eqs.~(\ref{R2D}) and (\ref{D2D}) and their analogs for an in-plane electric field. This dependence is SO coupling specific. Singularities at the poles are cut-off by a level width.

The $\theta$ dependences of the EDSR transition matrix element for $\tilde{\mbox{\boldmath$E$}}\parallel{\hat{\bf z}}$ and $\tilde{\mbox{\boldmath$E$}}\perp{\hat{\bf z}}$ are significantly different.  First, for the perpendicular-to-plane polarization, the factor $\sin2\theta$ in Eq.~(\ref{R2D}) and the factor $\sin\theta$ in Eq.~(\ref{D2D}) nullify these matrix elements at $\theta=0$. There are no such factors in the matrix elements of Sec.~\ref{sec:inplane}. Therefore, EDSR can be observed only in a tilted magnetic field \mbox{\boldmath$B$} when $\tilde{\mbox{\boldmath$E$}}(t)\parallel{\hat{\bf z}}$, but an arbitrary orientation of \mbox{\boldmath$B$} can be used (including $\mbox{\boldmath$B$}\parallel{\hat{\bf z}}$) for an in-plane $\hat{\mbox{\boldmath$E$}}(t)$. 

The second difference in EDSR with different $\tilde{\mbox{\boldmath$E$}}(t)$ polarizations concerns its intensity. This difference is essentially pronounced in the strong confinement regime, $\omega_0\gg\omega_s,\omega_c$. Comparing Eqs.~(\ref{R2D}) and (\ref{D2D}) with Eqs.~(\ref{APLR}) and (\ref{APLD}), one can estimate the corresponding matrix elements as $\sim \alpha_{R,D}\omega_s/\hbar\omega_0^2$ for $\tilde{\mbox{\boldmath$E$}}(t)\parallel{\hat{\bf z}}$ and as $\sim\alpha_{R,D}/\hbar\omega_c$ for $\tilde{\mbox{\boldmath$E$}}(t)\perp{\hat{\bf z}}$, respectively, when the inequality $\omega_c\agt|\omega_s|$ is satisfied. The first estimate indicates that when $\tilde{\mbox{\boldmath$E$}}(t)\parallel{\hat{\bf z}}$, EDSR is possible only due to the deviation of the system from the strict 2D limit. On the contrary, when $\tilde{\mbox{\boldmath$E$}}(t)\perp{\hat{\bf z}}$, EDSR survives in the strict 2D limit. The ratio of the matrix elements is about $\omega_c\omega_s/\omega_0^2\ll1$, hence, in the strong confinement limit an in-plane field $\tilde{\mbox{\boldmath$E$}}(t)$ is much more efficient than a perpendicular-to-plane field.  

For SIA, it is also instructive to compare the magnitudes of the EDSR transition matrix elements in a strong 2D confinement limit, $L_D^\psi$, described by  Eq.~(\ref{APLD}), and in the bulk, $L_{3D}\sim\delta m/\hbar^2$, as found by Rashba and Sheka \cite{RS61}.  We find $L_D^\psi/L\sim\omega_0/\omega_c\gg1$ using $\alpha_D\sim\delta m\omega_0/\hbar$ from Eq.~(\ref{alpha*}) below. This enhancement of EDSR in a QW in in-plane geometry can be attributed to a strong confinement with the square of the confinement momentum, $m\omega_0/\hbar$, large compared with $\lambda^{-2}$. Extraordinarily high efficiency of EDSR in the in-plane geometry has been emphasized in Ref.~[\onlinecite{APL04}].

We note that despite the fact that the intensity of EDSR with $\tilde{\mbox{\boldmath$E$}}(t)\parallel{\hat{\bf z}}$ is much less than with $\tilde{\mbox{\boldmath$E$}}(t)\perp{\hat{\bf z}}$, it is usually high enough for efficient electrically manipulating electron spins. Specific estimates can be found in Ref.~[\onlinecite{RE03}] and will be given in Sec.~\ref{discussion} below.

As  has been discussed above, there is a similarity in some of the properties of EDSR caused by BIA and SIA. The most striking difference in the effect of these mechanisms is seen in the angular dependences of the  EDSR intensity, especially in its dependence on the azimuth $\phi$, as will be discussed in Sec.~\ref{sec:Azimuth}. 

\subsubsection{Dependence of EDSR on the polar angle}
\label{sec:PolarAngle}

Using the equations of Sec.~\ref{sec:perp}, we will provide and discuss here the dependence of the EDSR intensity on the polar angle $\theta$ as applied to InAs. We restrict ourselves to SIA because Eq.~(\ref{eqDr}) for 2D Dresselhaus coupling is applicable only under strong confinement conditions when $\omega_0\gg\omega_c$; cf. Sec.~\ref{sec:inplane}. A theory of BIA controlled EDSR will be discussed in Sec.~\ref{sec:3D} in the framework of a more general approach.

It is seen from Fig.~\ref{fig:spectrum}b that the energy spectrum is rather different for $\omega_c<\omega_0$ and $\omega_c>\omega_0$. Therefore, these two cases will be considered separately.

The angular dependence of the EDSR intensity is controlled by the dependence of the square of the matrix element $L_R^z$ on $\theta$, Eq.~(\ref{R2D}), and by the population difference of the two spin sublevels that depends on the filling factor $\nu(\theta)=n_e/n_L(\theta)$, $n_e$ being the concentration of 2D electrons. To cut-off $L_R^z(\theta)$ near its pole and to find a realistic estimate of EDSR intensity {\it vs} $\theta$, we introduce a phenomenological level width $\Gamma$.

When $\omega_c\gg\omega_s$, one can consider a single spin-split level $E_0\pm\hbar\omega_s/2$ with $0<\nu<2$. For a Lorentzian level shape, the difference $\Delta\nu$ of the filling factors of two sublevels equals
\bea
\Delta\nu&=&{1\over\pi}\left[\arctan{{(\eta-E_0)+\hbar\omega_s/2}\over\Gamma}\right.\nonumber\\
&-&\left.\arctan{{(\eta-E_0)-\hbar\omega_s/2}\over\Gamma}\right],
\label{DeltaNu}
\eea
where $\eta(\nu)$ is the chemical potential that can be found from the equation
\bea
{{\eta(\nu)-E_0}\over\Gamma}&=&-{1\over{\tan\pi(\nu-1)}}+{\rm sign}\{\nu-1\}\nonumber\\
&\times&\sqrt{{1\over{\sin^2\pi(\nu-1)}}+\biggl({{\hbar\omega_s}\over{2\Gamma}}\biggr)^2}.
\label{ChPot}
\eea
The dependence of $\Delta\nu$ on $\nu$ is shown in the Fig.~\ref{fig:TwoLevel}. For small $\Gamma$, this is a triangle. Its vertices can be considered as cusps of the $\Delta\nu$ {\it vs} $\nu$ curve. An up-cusp appears at $\nu=1$, and two down-cusps at $\nu=0$ and 2. Note, the up-cusp appears at odd $\nu$, while down-cusps at even $\nu$ values. With increasing $\Gamma$, the cusps are smeared, however, they manifest themselves as a well pronounced maximum and two minima even for $\Gamma$ as large as $\Gamma=\hbar\omega_s/2$. Below, we will find similar patters in the spectra of multi-level systems that are well pronounced even for a strongly tilted field \mbox{\boldmath$B$} when $\omega_c\cos\theta\sim\omega_s$. However, because of level crossings the regular alternation of up- and down cusps (and their correspondence to odd and even $\nu$ factors, respectively) can be violated; an example of such an anomaly can be seen in Fig.~\ref{fig:Polar}b. 

In what follows, we use a Gaussian level shape $\exp[-(E-E_0)^2/\Gamma^2]$ because it provides pronounced features in the intensity distribution that can be reliably assigned. This procedure will be applied to the denominator ${\cal D}(\theta)$ of Eq.~(\ref{R2D}) where it cuts the pole and also when calculating the populations of different orbital levels and their spin sublevels. However, we do not renormalize the frequencies $\omega_c$ and $\omega_s$ in the numerator of Eq.~(\ref{R2D}), and a similar procedure will be applied everywhere below. 

In Fig.~\ref{fig:Polar}a is shown the dependence of EDSR intensity on $\theta$ for $\omega_c=\omega_0/2$ and the electron concentration $n_e=2n_L(\theta=0)$. Therefore, for $\theta=0$ both spin sublevels of the $n_\xi=0, n_\eta=0$ level are filled, and EDSR develops only when $\mbox{\boldmath$B$}$ becomes tilted, $n_L(\theta$) decreases, and the $n_\xi\geq 1, n_\eta=0$ level is getting occupied. The general shape of the curve is dominated by a pronounced maximum that is achieved near the pole of the denominator, $\omega_\xi=|\omega_s|$, and shifts to lower angles with increasing $\Gamma$. Strong suppression of the EDSR intensity for small angles and for $\theta\approx\pi/2$ originates from the factor $\sin2\theta$ in Eq.~(\ref{R2D}).

Assigning specific features in Fig.~\ref{fig:Polar}a can be done by following the populations of successive levels and the behavior of the chemical potential $\eta$ with increasing $\theta$; $\eta(\theta)$ is shown in the insert. For small $\Gamma$ values, populating $n_\xi$ level begins only after both spin components of the previous level, with the quantum number $(n_\xi-1)$, are completely occupied. This regularity holds up to $\nu=10$ and manifests itself in regular tooth-like pattern of the curve $\eta(\nu)$ for small level width, $\Gamma/\hbar\omega_0=0.01$. The same regularity can also be seen from the $\theta$-dependence of the EDSR intensity for $\Gamma/\hbar\omega_0=0.05$. Similarly to Fig.~\ref{fig:TwoLevel}, up-cusps correspond to odd filling factors $\nu$, while down-cusps to even $\nu$. The larger is the gap between the successive energy levels, the larger is the ``jump" in $\eta(\nu)$, and the more pronounced is the corresponding cusp. Small shifts of the cusps from integer $\nu$-values originate from the final level width $\Gamma$. Because of the general growth of the EDSR intensity with $\theta$ at the left slope of the principal maximum, up-cusps at this slope are shifted to the right from odd $\nu$-values while down-cusps are shifted to left from even $\nu$-values. Remarkably, all these features are distinctly seen in the intensity pattern for $\Gamma/\hbar\omega_0=0.05$ despite the fact that the function $\eta(\nu)$ is already rather smooth for this level width.

We have calculated only transition intensities and did not calculate the corrections to the $g$-factor that originate from the  band nonparabolicity and SO coupling. These corrections should depend on the quantum numbers $n_\xi,n_\eta$. Because of the regular level alternation described above, only a single spin-flip frequency should be seen inside each window confined between two successive even values of $\nu$; $n_\xi$ values for specific windows are indicated in the insert to Fig. 3a. Two spin-flip frequencies can be seen simultaneously only inside narrow regions of $\theta$ where $\nu$ passes through  even values.

The EDSR intensity for $\omega_c/\omega_0=2$ and the electron concentration $n_e=2n_L(\theta=0)$ is shown in Fig.~\ref{fig:Polar}b; in this case, $\omega_\eta$ is the lower spectrum branch. The basic shape of the spectrum is the same: it is dominated by a strong maximum at $\omega_\eta\approx |\omega_s|$. However, there is a considerable difference in the fine structure. First, because the principal maximum is achieved at a lesser value of $\theta$, fine structure is distinctly seen on both sides of the maximum. Second, level intersections arise at relatively small quantum numbers $n_\eta$; the intersection of $(1,\downarrow)$ and $(2,\uparrow)$ levels is the first one. However, of much more importance is the intersection of $(2,\downarrow)$ and $(3,\uparrow)$ levels because Fermi energy $\eta$ passes through the intersection point. As a result, the simple regularity in populating successive levels does not hold any more, and alternation of up- and down-cusps is violated. Between the well pronounced $\nu=5$ up-cusp and the $\nu=7$ down-cusp there exists a smeared $\nu=6$ up-cusp. For this reason, signs of the following cusps change: a strong odd-$\nu$ cusp $\nu=7$ turns into a down-cusp, while an even-$\nu$ cusp $\nu=8$ becomes an up-cusp. 

The intersection of $(2,\downarrow)$ and $(3,\uparrow)$ levels happens when $3\omega_\eta-|\omega_s|/2=2\omega_\eta+|\omega_s|/2$, i.e., it coincides with the zero of the denominator of Eq.~(\ref{R2D}) and the principal peak of EDSR. Because this intersection point also coincides with the Fermi level, to the left from this point only the $(2,\uparrow)\rightarrow(2,\downarrow)$ transition to the half-populated $(2,\downarrow)$ level is allowed. On the contrary, to the right from the intersection point both the $(2,\uparrow)\rightarrow(2,\downarrow)$ transition to the empty $(2,\downarrow)$ level and the $(3,\uparrow)\rightarrow(3,\downarrow)$ transition to the half-populated $(3,\downarrow)$ level are allowed, hence, the intensity of EDSR is expected to increase abruptly by a factor of three. The inflection point seen between the $\nu=5$ cusp and the maximum of the curve reflects this discontinuity smeared by the level width $\Gamma$; smearing is rather strong because the slopes of $(3,\uparrow)$ and $(2,\downarrow)$ levels plotted {\it vs} $\theta$ are very close. The asymmetry of the peak with respect to the $\omega_\eta=|\omega_s|$ point, including shifting its maximum to the right from this point, is a different manifestation of the effect of the increase in the number of electronic channels at the intersection point. Also, the coexistence of two transition channels, $(2,\uparrow)\rightarrow(2,\downarrow)$ and $(3,\uparrow)\rightarrow(3,\downarrow)$, manifests itself in the high EDSR intensity of the background over which the even-$\nu$ cusp at $\nu=6$ is hardly seen.

The coexistence regions of different transitions are shown in the insert to Fig.~\ref{fig:Polar}b. Transitions between spin sublevels of the levels $n_\eta=2$ and $n_\eta=3$, coexisting in a wide region of filling factors, should manifest themselves through splitting the spin-flip line into two components with close $g$-factors. The same is valid for $n_\eta=3$ and $n_\eta=4$ levels (for larger $\theta$ values).

All the above analysis is based on a phenomenological approach with a single level width parameter $\Gamma$. In the framework of such a theory, it describes both the widening of $\xi$ and $\eta$ energy levels and the width of the spin-flip transition lines. However, the real physics might be rather different. It has been shown by Mel'nikov and Rashba\cite{MR71} that impurity scattering results in a dramatic narrowing of spin-flip lines, and narrow EDSR lines have been observed in dirty semiconductor materials by Bell long before.\cite{B62} The underlying mechanism is the dynamical narrowing that is also responsible for the D'yakonov-Perel' spin relaxation mechanism.\cite{DP71} Therefore, when two or more wide $\xi$ or $\eta$ levels are partially populated, one should anticipate observing several narrow spin-flip lines in the EDSR spectrum. Indeed, two narrow spin-flip lines were observed in $n$-type inversion layers on InSb, one of them showing a Fano-type profile;\cite{Merkt} apparently, they were caused by EDSR, but the mechanism has not been specified. The detailed theory of such spectra should be model dependent.

Therefore, EDSR spectra include rich and nontrivial information about energy spectra. Extracting it can be achieved by a detailed analysis of the spectra supported by calculating energy levels and their populations. 

\subsubsection{Azimuth dependence of EDSR intensity}
\label{sec:Azimuth}

In the previous section, we considered a perpendicular-to-plane electric field, ${\tilde{\mbox{\boldmath$E$}}}\parallel{\hat{\bf z}}$, concentrated on the details of the EDSR dependence on the polar angle $\theta$, and restricted ourselves with SIA. Because of the axial symmetry of the Hamiltonian ${\hat H}_R$, the azimuth dependence of EDSR intensity was isotropic.
 
In this section we provide the full angular dependence of EDSR intensity with a special emphasis on its azimuth dependence; and with this end in view we consider both basic polarizations of the ac field $\tilde{\mbox{\boldmath$E$}}$ and both 2D Hamiltonians, ${\hat H}_D$ and ${\hat H}_R$. Because the equations of Sec.~\ref{sec:inplane} derived for the Hamiltonian ${\hat H}_D$ are applicable only under the conditions of strong confinement, we accept that $\omega_0\gg\omega_c,\omega_s$, and that the electron concentration is low enough, hence, electrons populate only the lowest quantization level, $n_\xi, n_\eta=0$, for all polar angles $\theta$ of interest; this restriction will be removed in the more general theory of Sec.~\ref{sec:3D}. To suppress fine features of the $\theta$ dependence, we choose a relatively large level width, $\Gamma=0.2\omega_0$; it suppresses the resonance at $\omega_\zeta=|\omega_s|$ and enhances EDSR in the small $\theta$ region.

At first we keep ${\tilde{\mbox{\boldmath$E$}}}\parallel{\hat{\bf z}}$ and consider the $\phi$ dependence originating from the Hamiltonian ${\hat H}_D$. It possesses only the two-fold symmetry of the ${\mbox{\boldmath$C$}}_{2v}$ group. However, the matrix element $L^z_D$ of Eq.~(\ref{D2D}) is odd to $\pi/2$ rotations. Therefore, the intensity of EDSR acquires the four-fold symmetry that manifests itself in Fig.~\ref{fig:AzimuthPerp}a. However, the joint effect of ${\hat H}_D$ and ${\hat H}_R$ (the latter one is totally symmetric!) eliminates this additional symmetry. The effect of the interference of BIA and SIA is seen in Fig.~\ref{fig:AzimuthPerp}b; it is especially strong for $\alpha_D=\pm\alpha_R$. The degree of asymmetry should allow measuring the ratio $\alpha_D/\alpha_R$.

Now we turn to the in-plane geometry, ${\tilde{\mbox{\boldmath$E$}}}\perp{\hat{\bf z}}$, where the symmetry of the indicatrix is additionally lowered because the azimuth $\psi$ of the field $\tilde{\mbox{\boldmath$E$}}$ establishes a new preferred direction in the $(x,y)$ plane. Figures \ref{fig:InPlane}a and \ref{fig:InPlane}b are based on Eq.~(\ref{APLR}). Because of the isotropy of the Hamiltonian ${\hat H}_R$, the EDSR intensity depends only on the difference of the azimuths $\psi$ and $\phi$ of the electric field ${\tilde{\mbox{\boldmath$E$}}}$ and magnetic field ${\mbox{\boldmath$B$}}$, respectively. Therefore, for $\phi=\psi$, the EDSR intensity does not depend on the azimuth as it is reflected in the rotational symmetry of Fig.~\ref{fig:InPlane}a. Remarkably, in this geometry Eq.~(\ref{APLR}) reduces to its first term that vanishes when $\omega_c^2\cos^2\theta+\omega_s=0$. That is why the angular indicatrix of Fig.~\ref{fig:InPlane}a, drawn for InAs with $\omega_s=-0.17\omega_c$, consists of two sheets touching in a single point. The low-$\theta$ feature became visible because of the large $\Gamma=0.2\omega_0$. From this standpoint, it is instructive to compare the figures from our previous paper, Ref.~\onlinecite{APL04}, with the set of figures of this section; they complement each other rather well. In Fig.~\ref{fig:InPlane}b, the direction of electric field is fixed along the [110] crystallographic axis, and the $\phi$ dependence of the EDSR intensity shows only the two-fold axis symmetry.

Figures \ref{fig:InPlane}c and \ref{fig:InPlane}d are based on Eq.~(\ref{APLD}). 
For Fig.~\ref{fig:InPlane}c, ${\tilde{\mbox{\boldmath$E$}}}\parallel{\mbox{\boldmath$B$}}_\perp$  and therefore $\phi=\psi$, ${\mbox{\boldmath$B$}}_\perp$ being the projection of ${\mbox{\boldmath$B$}}$ onto the confinement plane. The matrix element of Eq.~(\ref{APLD}) has only a two-fold symmetry axis that corresponds to the symmetry of the Hamiltonian ${\hat H}_D$. However, its square possesses four-fold symmetry that one easily recognizes in Fig.~\ref{fig:InPlane}c. In Fig.~\ref{fig:InPlane}d, direction of ${\tilde{\mbox{\boldmath$E$}}}$ is fixed as ${\tilde{\mbox{\boldmath$E$}}}\parallel[110]$, and the symmetry of EDSR intensity is reduced again to a two-fold axis.

Figure \ref{fig:InPlane}e illustrates the joint effect of ${\hat H}_R$ and ${\hat H}_D$ Hamiltonians for ${\hat{\mbox{\boldmath$E$}}}\parallel{\mbox{\boldmath$B$}}_\perp$. This figure, drawn for $\alpha_R=\alpha_D$, reflects the two-fold symmetry inherent in the BIA Hamiltonian ${\hat H}_D$. 

\subsection{Experimental data}
\label{sec:exper}

EDSR is well documented in 3D where various mechanisms of it have been discovered and identified. \cite{B62,MBK,Dobr84,RS91} Experimental data regarding 2D systems are very scarce yet. Ironically, in 2D the effect of spin-orbit coupling on spin resonance is best understood in Si/SiGe quantum wells where SO coupling is notoriously weak;\cite{JWSMS02,TLJS05} the success is mostly based on long spin coherence times. Electrical monitoring of electron spins in AlGaAs parabolic quantum wells near the $g\approx0$ point has been achieved by the $\hat g$-tensor modulation technique.\cite{KMDGLA03} It was only very recently that EDSR driven by the orbital mechanism has been reported for a A$_3$B$_5$ quantum well.\cite{SLDD05} And ironically again, this observation has been made with an AlAs quantum well where the weakness of SO coupling manifested itself in a small $g$-shift, $g\approx1.99$. Nevertheless, the intensity of EDSR exceeded the intensity of EPR (excited by the magnetic component of microwave field) by four orders of magnitude. As Schulte {\it et al.} emphasize in their paper,\cite{SLDD05} the very fact of the observation of a spin flip line represented a puzzle because in their sample the EPR intensity was two orders of magnitude smaller than the noise level.

We note that AlAs is not a propitious material for applying our theory because of the small effective mass, $m/m_0=0.46$, and small $g$-factor, $g=1.99$. As a result, even for magnetic fields $B\sim1$ T, the cyclotron and spin frequencies are in the microwave range and the criterion of strong cyclotron quantization is at the verge of its applicability. In Ref.~[\onlinecite{SLDD05}], $f_c\tau_p=2.4$ and $f_s\tau_p=1.1$ for $B=1.219$ T, and $f_c\tau_p=0.67$ and $f_s\tau_p=0.31$ for $B=0.3349$ T, where $f_{c,s}=\omega_{c,s}/2\pi$ and $\tau_p$ is the momentum relaxation time. Apparently for this reason, Ref.~[\onlinecite{SLDD05}] does not include any data regarding cyclotron absorption, and all the discussion is provided in terms of non-quantized orbital dynamics. The basic conclusions of Schulte {\it et al.}\cite{SLDD05} are as follows. The signal of spin resonance originates from the effective magnetic field caused by the SO interaction. The polarization dependence of the resonance is well described by the effective field $\mbox{\boldmath$B$}_{\rm eff}=(2\alpha_R/g\mu_B)(\mbox{\boldmath$k$}\times{\hat{\bf z}})$ following from the Hamiltonian of Eq.~(\ref{eqR}), and its intensity suggests $\alpha_R\approx5\times10^{-12}$ eV cm as a crude estimate for the SO coupling constant.

The high intensity of EDSR in the ${\mbox{\boldmath$B$}}\parallel{\hat{\bf z}}$ geometry, even when the sample was positioned close to the node of the microwave electric field $\tilde{\mbox{\boldmath$E$}}$, was observed only when the field $\tilde{\mbox{\boldmath$E$}}$ was in-plane polarized; unfortunately, EDSR in a tilted magnetic field has not been studied. EDSR has not been seen with $\tilde{\mbox{\boldmath$E$}}\parallel{\hat{\bf z}}$.\cite{SLDD05} These observations are in agreement with the predictions of Refs.~[\onlinecite{RE03}] and [\onlinecite{APL04}] and the conclusions of Sec.~\ref{PolarDepen}.

\section{3D Dresselhaus Hamiltonian}
\label{sec:3D}

The physical parameter that allows reducing the Hamiltonian ${\hat{\cal H}}_D$ of Eq.~(\ref{eq11}) to its 2D form ${\hat H}_D$ is a small confinement length $d$ or, what is the same, a large confinement frequency $\omega_0$. Therefore, the expressions found in Sec.~\ref{EDSR} for the 2D Dresselhaus Hamiltonian of Eq.~(\ref{eqDr}) and arbitrary values of $\omega_c/\omega_0$ and $(n_\xi,n_\eta)$ are mostly of methodical interest. They have provided, side by side with equations found for the Rashba Hamiltonian of Eq.~(\ref{eqR}), an important outlook on the comparative strength of EDSR excited by in-plane and perpendicular-to-plane electric fields $\tilde{\mbox{\boldmath$E$}}(t)$. However, a consistent description of 3D Dresselhaus systems can be achieved only by using the Hamiltonian ${\hat{\cal H}}_D$. Technically, calculating matrix elements of EDSR for that Hamiltonian is a challenging task. We will show in this section, as applied to a field $\tilde{\mbox{\boldmath$E$}}(t)\parallel{\hat{\bf z}}$, that using the operators of Sec. \ref{operators} in conjunction with the identities of Appendix \ref{identities} is a powerful tool that allows solving the problem and deriving explicit expressions for matrix elements.

In the primed reference frame, the term ${\hat{\cal H}}_D$ acquires a form $\hat{\cal H}_D=\delta\sigma_{j'}\hat{\kappa}_{j'}$ with $\hat{\kappa}_{j'}=U_{jj'}\hat{\kappa}_{j}$, because the matrix $U$ is orthogonal, hence, $U^{-1}=U^T$. Of all the multitude of terms that enter in this expression, we need to select only those that contribute to the matrix element of Eq.~(\ref{transition}). First, the selection rules in the spin operators show that
 \be
\langle \uparrow|\hat{\cal H}_{D}|\downarrow\rangle=\delta\hat{\cal K}=\delta(\hat{\kappa}'_{x'}-i\hat{\kappa}'_{y'})~.
\label{spinflip}
\ee
Second, because the operators of coordinates are linear in $(a_\zeta,a_\zeta^+)$, the selection rules in the orbital operators in equations similar to Eq.~(\ref{transition}) select only those parts of the operators ${\hat{\kappa}'}_{j'}$ that change one of the quantum numbers, either $n_\xi$ or $n_\eta$, by $\pm1$. Therefore, in what follows we denote by ${\hat{\kappa}'}_{j'}$ these parts of the operators. Straightforward calculations in the crystal coordinate frame result in the following expressions for them
\bea
\hat{\kappa}'_{j}&=&\sum_{\zeta}[C_{j;\zeta\zeta}\hat{a}_\zeta\hat{n}_\zeta+C^*_{j;\zeta\zeta}\hat{n}_\zeta\hat{a}^+_\zeta]\nonumber\\
&+&\sum_{\zeta'\neq\zeta}(C_{j;\zeta\zeta'}\hat{a}_\zeta+C^*_{j;\zeta\zeta'}\hat{a}^+_\zeta)(2\hat{n}_{\zeta'}+1)~,
\eea
where $\hat{n}_\zeta=\hat{a}^+_\zeta\hat{a}_\zeta$. Coefficients $C_{x;\zeta\zeta'}$ read as
\bea
C_{x;\zeta\zeta'}&=&Y_\zeta(X_{\zeta'} Y_{\zeta'}^*+Y_{\zeta'} X_{\zeta'}^*)-Z_\zeta(X_{\zeta'} Z_{\zeta'}^*+Z_{\zeta'}X_{\zeta'}^*)\nonumber\\
&+&X_\zeta(|Y_{\zeta'}|^2-|Z_{\zeta'}|^2)~, 
\label{Ds}
\eea
and coefficients $C_{y;\zeta\zeta'}$ and $C_{y;\zeta\zeta'}$ can be found from $C_{x;\zeta\zeta'}$ by cyclic permutatins of $X$, $Y$, and $Z$ factors in the right hand side of Eq.~(\ref{Ds}). In these notations,
\bea
\hat{\cal K}&=&\sum_{\zeta}[{\cal C}_{\zeta\zeta}^-\hat{a}_\zeta\hat{n}_\zeta+{\cal C}^+_{\zeta\zeta}\hat{n}_\zeta\hat{a}^+_\zeta]\nonumber\\
&+&\sum_{\zeta'\neq\zeta}({\cal C}^-_{\zeta\zeta'}\hat{a}_\zeta+{\cal C}^+_{\zeta\zeta'}\hat{a}^+_\zeta)(2\hat{n}_{\zeta'}+1)~
\label{Koper}
\eea
with  ${\cal C}^{\pm}_{\zeta\zeta'}$
\bea
{\cal C}^{+}_{\zeta\zeta'}&=&(U_{jx'}-iU_{jy'})C_{j;\zeta\zeta'}^*~,\nonumber\\
{\cal C}^{-}_{\zeta\zeta'}&=&(U_{jx'}-iU_{jy'})C_{j;\zeta\zeta'}~.
\label{Ccoefs} 
\eea 

To calculate the probability of the spin-flip transition caused by a perpendicular-to-plane field $\mbox{\boldmath $\tilde{E}$}(t)\parallel{\bf\hat{z}}$, we express the $z$-coordinate operator in terms of operators $(\hat{a}_\zeta,\hat{a}_\eta^+)$
\be
\hat{z}=\sum_{\zeta=\xi,\eta}~l_{z\zeta}(\hat{a}_\zeta+\hat{a}^+_\zeta),
\label{z1}
\ee
where $l_{z\zeta}$ are defined in Eq.~(\ref{ls}). Substituting (\ref{Koper}) and (\ref{z1}) into Eqs.~(\ref{Tmatrix})  and (\ref{transition}), one arrives at
\bea
L_z(n_\xi,n_\eta)&=&
\langle n_\xi,n_\eta,\uparrow\vert\hat{z}_{\rm so}\vert n_\xi,n_\eta,\downarrow\rangle \nonumber\\
&=& -{\delta\over\hbar}
\sum_{\zeta\zeta'\nu}(2n_{\zeta'}+1){{\ell_{z\zeta}{\cal C}^\nu_{\zeta\zeta'}}\over{\omega_\zeta+\nu\omega_s}}
\label{z2}
\eea
 with $\nu=\pm$. Remarkably, despite the fact that diagonal and nondiagonal coefficients ${\cal C}_{\zeta\zeta'}$ enter into $\hat{\cal K}$ in a nonsymmetrical way, in (\ref{z2}) the symmetry is restored, and $\zeta$ and $\zeta'$ take both values, $\zeta,\zeta'=\xi,\eta$. Equation (\ref{z2}) can be conveniently rewritten as
\bea
L_z(n_\xi,n_\eta)
&=&-{\delta\over \hbar}\sum_{\zeta\zeta'} l_{z\zeta}(2n_{\zeta'}+1)\nonumber \\
&\times&\left(\omega_\zeta {{\cal C}^{-}_{\zeta\zeta'}+{\cal C}^{+}_{\zeta\zeta'} \over \omega_\zeta^2-\omega_s^2} +\omega_s{{\cal C}^{-}_{\zeta\zeta'}-{\cal C}^{+}_{\zeta\zeta'} \over \omega_\zeta^2-\omega_s^2}\right).
\label{3DTR}
\eea

It is important for following calculations that the dependence of coefficients $C_{j;\zeta\zeta'}$ on indices $\zeta$ and $\zeta'$ can be factorized (see Appendix \ref{coeffb}) as
\be
C_{j;\zeta\zeta'}=\beta_{\zeta'}[d_jR_{j\zeta}+(\mbox{\boldmath$R$}_\zeta\times\mbox{\boldmath$D$})_j]~
\ee
with
\be
\beta_\xi=-m\omega_\xi/2\hbar,\,\,\beta_\eta=m\omega_\eta/2\hbar.
\label{betas}
\ee
Therefore, the dependence of the coefficients $C_{j;\zeta\zeta'}$ on the last subscript is universal and rather trivial. Their dependence on the two first subscripts and the angles $(\theta,\phi)$ can be written in terms of three 3D vectors
\bea
\mbox{\boldmath$R$}_\zeta&=&(X_\zeta,Y_\zeta,Z_\zeta),~\mbox{\boldmath$d$}=(b_{Y-Z},b_{Z-X},b_{X-Y}),\nonumber\\
\mbox{\boldmath$D$}&=&(b_{YZ},b_{ZX},b_{XY}),
\eea 
of which the first one is defined by Eqs.~(\ref{eqn15}), while $\mbox{\boldmath$d$}$ and $\mbox{\boldmath$D$}$ depend only on $(\theta,\phi)$. This structure of coefficients $C_{j;\zeta\zeta'}$ can be established by direct calculations. Formulas for coefficients $b$ with different indices are given by Eq.~(\ref{bms}). 

Details related to calculating matrix elements of Eq.~(\ref{3DTR}) are described in Appendix \ref{coeffb}. We present the final expression following from Eqs.~(\ref{z3}) in the form that is similar to Eq.~(\ref{D2D}) to simplify the comparison with the results of Sec.~\ref{sec:perp}
\bea
L_z(n_\xi,n_\eta)&=&-{{\alpha_D^{\rm eff}(n_\xi,n_\eta)\omega_s}\over{\hbar}}
{{\omega_c\sin\theta}\over{{\cal D}(\theta)}}{\rm sign}\{\omega_0-\omega_c\}\nonumber\\
&\times&[\sin2\phi\cos\theta~F_s-i\cos2\phi~F_c]~,
\label{Answer}
\eea
where
\bea
\alpha_D^{\rm eff}(n_\xi,n_\eta)&=&\alpha_D^*\omega_0[\omega_\eta(1+2n_\eta)-\omega_\xi(1+2n_\xi)]/\Omega^2(\theta)\nonumber\\ 
 {\rm with}~~~~\alpha_D^*&=&-{{\delta m\omega_0}/{2\hbar}}.
\label{alpha*}
\eea
Functions $F_s$ and $F_c$ are defined as
\bea
F_s(\theta)=(\omega_c-\omega_s)&+&{1\over{\omega_0^2}}\left[2\omega_c\omega_s^2-3\omega_c^3{2\cos^2\theta-\sin^2\theta\over 2}\right.\nonumber\\
&+&\left.\omega^2_c\omega_s{1+3\sin^2\theta\over2}\right]~,
\label{Fs}
\eea
\bea
F_c(\theta)=(\omega_c\cos^2\theta-\omega_s)&+&{1\over{\omega_0^2}}[\omega_c\omega_s^2(1+\cos^2\theta)\nonumber\\
-\omega_c^3\cos^2\theta&+&\omega^2_c\omega_s\cos2\theta]~.
\label{Fc}
\eea
It is seen from Eq.~(\ref{alpha*}) that the effective coupling constant $\alpha_D^{\rm eff}$ depends on the population of the both  $\xi$ and $\eta$ -- type energy levels. This factor reduces to a constant when (i) only the lower level of the spatial quantization is populated, $n_\eta=0$, and (ii) the Fermi energy is small compared to the spatial quantization energy, $n_\xi\omega_\xi\ll\omega_\eta$. In the strong confinement limit, with $\Omega^2\approx\omega_0^2$, we get $\alpha_D^{\rm eff}\approx-\delta m\omega_0/2\hbar$. Having in mind that $m\omega_0/2\hbar\approx(\pi/d)^2$, $d$ playing the role of the confinement length, we arrive at the usual estimate for $\alpha_D$. However, when the criterion of strong spatial quantization is not fulfilled, $\alpha_D^{\rm eff}$ influences the dependence of the EDSR intensity on the polar angle of \mbox{\boldmath$B$}.

Functions $F_s$ and $F_c$ are defined in such a way that their first terms dominate in the strong confinement limit, $\omega_0\gg\omega_c,\omega_s$, and these coefficients coincide exactly with the corresponding coefficients in Eq.~(\ref{D2D}). Therefore, in this case Eq.~(\ref{Answer}) reduces to Eq.~(\ref{D2D}). However, it will be shown in the next section that when $\omega_0\sim\omega_c,\omega_s$, the factors $F_s$ and $F_c$ influence profoundly the intensity of EDSR.

\subsection{Angular dependence of EDSR}
\label{sec:AngDepen}

In Fig.~\ref{fig:Polar}a,b the dependence of the EDSR intensity on the polar angle $\theta$ has been presented for the Hamiltonian ${\hat H}_R$ describing SIA. It has been discussed at the end of Sec.~\ref{sec:perp} that the applicability of the Hamiltonian ${\hat H}_D$ is more restricted as compared to ${\hat H}_R$. Therefore, for calculating  the angular dependence of EDSR intensity coming from BIA in a wide range of polar angles $\theta$, we apply the general results derived in the previous section. Comparison of Eqs.~(\ref{D2D}) and (\ref{Answer}) shows that this generalization influences EDSR intensity in three ways. First, through new terms in functions $F_s(\theta)$ and $F_c(\theta)$. Second, through $\theta$ dependence of $\Omega(\theta)$. Third, the transition probability acquires a dependence on population numbers $n_\xi,n_\eta$ entering into the effective coupling constant $\alpha_D^{\rm eff}(n_\xi,n_\eta)$. Also, as distinct from ${\hat H}_R$, the Hamiltonian ${\hat{\cal H}}_D$ does not possess continuous rotational symmetry about the $z$ axis. Therefore, we calculate the $\theta$-dependence of the EDSR intensity for two specific values of $\phi$, $\phi=0$ and $\phi=\pi/4$.

The polar angle dependences of the EDSR intensities presented in Fig.~\ref{fig:BIA} have been calculated for the same values of $\omega_c/\omega_0$ and $\omega_s/\omega_0$ as the data of Fig.~\ref{fig:Polar} and $n_e=2n_L(\theta=0)$. We have also used the same Gaussian level broadening. Because we disregarded the effect of SO coupling on the energy spectrum, all the above discussion of the energy spectrum and the $\theta$-dependence of the chemical potential $\eta(\theta)$ is completely applicable to the current case. Hence, the basic behavior of the intensity including the principal maximum near $\omega_\zeta=|\omega_s|$ and existence of a system of up- and down-cusps remains intact. The difference comes from the envelope factors $F_s(\theta)$ and $F_c(\theta)$, from the $\theta$-dependence of $\alpha_D^{\rm eff}$, and from the fact that for BIA the matrix element of EDSR does not vanish when $\theta\rightarrow\pi/2$. For large $\theta$ the cyclotron frequency $\omega_c(\theta)=\omega_c\cos\theta$  vanishes and our theory becomes inapplicable. Therefore, we restrict ourselves with the region $\theta\alt0.9\times\pi/2$.

In Fig.~\ref{fig:BIA}a, the EDSR intensity is displayed for $\phi=0$. It shows cusps similar to Fig.~\ref{fig:Polar}a, however, all features seem small in the scale of the figure. This difference in the scale can be understood from the shape of the envelope function shown in the inset. It includes the factors from Eq.~(\ref{Answer}) that depend on $\theta$ explicitly and do not involve quantum numbers $n_\xi,n_\eta$. As distinct from the envelope function of Fig.~\ref{fig:Polar}a that includes a factor $\sin^22\theta$, Eq.~(\ref{R2D}), this envelope function changes slowly on the right from the principal maximum at $\omega_\xi\approx|\omega_s|$ and does not vanish at $\theta=\pi/2$. Therefore, the EDSR intensity remains strong for large $\theta$, and we have shown a curve for a very small $\Gamma=0.02\hbar\omega_0$ to demonstrate the persistence of the principal maximum. At larger $\Gamma$ values the maximum washes out, and for $\omega_c\cos\theta\alt\Gamma/\hbar$ our approach is no more applicable. Remarkably, for $\omega_c=\omega_0/2$ the envelope function is close to its shape in the strong confinement limit as one can see from the inset.

In Fig.~\ref{fig:BIA}b, the intensity of EDSR is shown for the same value $\omega_c/\omega_0=0.5$ but for $\phi=\pi/4$. In this case the envelope function includes a factor $\sin^22\theta$ and vanishes for $\theta=\pi/2$. For this reason, Fig.~\ref{fig:BIA}b bears much more similarity with Fig.~\ref{fig:Polar}a. Deviation of the envelope function from its strong confinement limit is by a factor of about 2. Comparison of Figs.~\ref{fig:BIA}a and \ref{fig:BIA}b suggests a considerable azimuth dependence of EDSR intensity, especially in the large $\theta$ region.

It is very instructive to compare panels c and d of Fig.~\ref{fig:BIA} drawn for $\omega_c/\omega_0=2$ with the panels a and b of the same figure. The first difference is easily seen from the envelope functions shown in inserts. They show zeros for intermediate values of $\theta$, $\theta\approx0.82\times\pi/2$ and $\theta\approx0.61\times\pi/2$ for $\phi=0$ and $\phi=\pi/4$, respectively. The corresponding minima originate from the zeros of functions $F_c(\theta)$ and $F_s(\theta)$, respectively, appear only for specific values of the azimuth $\phi$, and strongly influence the EDSR intensity.\cite{zeros} The existence of an additional zero that is common for both panels c and d (but absent from a and b) and persists for arbitrary $\phi$ is the second difference. It originates from the zero of $\alpha_D^{\rm eff}(n_\xi,n_\eta)$ at $3\omega_\eta(\theta)=\omega_\xi(\theta)$ that is located at $\theta\approx0.42\times\pi/2$. For this $\theta$ value, the filling factor $\nu$ is within $2<\nu<3$. Hence, the spin flip transitions involve the $n_\xi=0,n_\eta=1$ electrons. The third important difference is a strong enhancement of EDSR intensity at $\omega_c=2\omega_0$ as compared with $\omega_c=\omega_0/2$. Intensities in Fig.~\ref{fig:BIA} are given in arbitrary units, but these units are the same for all panels when normalized on the confinement frequency $\omega_0$. The origin of the enhancement can be easily understood from Eq.~(\ref{D2D}) that in the strong confinement regime includes a factor $\alpha_D\omega_s/\omega_0^2$ with $\alpha_D\propto\omega_0$. Therefore, when $B$ increases, one can expect enhancement in EDSR intensity by a factor of about $(\omega_s/\omega_0)^2$.

\section{Discussion}
\label{discussion}

The results of our calculations allow one to evaluate the intensity of EDSR and to compare it with the intensity of the usual paramagnetic resonance (EPR) caused by a time dependent magnetic field. Their ratio depends on the SO coupling mechanism and the geometry of the experiment. In both the EDSR and EPR experiments, the efficiency is controlled by two major factors: (i) by the characteristic length
$l$ which in the case of EDSR is equal to the matrix element of the spin-flip transition of Eq. (\ref{transition}), and (ii) by the population difference of two spin sublevels that depends on the 2D electron concentration and the polar angle $\theta$ between the external magnetic field and direction of 2D confinement.

The effect of the latter factor does not depend on the mechanism of spin-flip transitions and influences in a similar way both the EDSR and EPR. It manifests itself in the modulation of the spin-flip transition intensity as a function of $\theta$ and depends on the population of spin sublevels controlled by the $\theta$ dependent position of the chemical potential $\eta(\theta)$. The corresponding up- and down-cusps are seen in Figs.~\ref{fig:Polar} and \ref{fig:BIA} showing the polar angle dependence of EDSR intensity for both basic SO coupling mechanisms. Level broadening washes out this population dependent fine structure. However, for a reasonably small broadening, $\Gamma/\hbar\omega_c \alt0.1$, this fine structure should allow studying the energy spectra of quantum wells. 

A geometry with the electric field $\tilde{\mbox{\boldmath$E$}}$  perpendicular to the confinement plane, similar to the one used by Kato {\it et al.}\cite{KMDGLA03}, in principle allows access to electron spins at a nanometer scale.  However, spin-flip transitions occur only in a tilted magnetic field and vanish when $\theta=0$, as is seen from Eqs.~(\ref{R2D}), (\ref{D2D}), and (\ref{Answer}). Indeed, it is a tilted magnetic field that couples the in-plane and perpendicular to the plane electron motions and allows a perpendicular to the plane field $\tilde{\mbox{\boldmath$E$}}$ to produce cyclotron and spin-flip transitions. 

The efficiency of the Dresselhaus mechanism of EDSR in this geometry can be evaluated by using the effective length $l_D^\perp=L_z(n_\xi=0,n_\eta=0)$ defined by Eq.~(\ref{Answer}). In the strong confinement limit,$\omega_0\gg\omega_c,\omega_s$, this length is about $l_D^\perp\sim(\alpha^*_D/\hbar\omega_0)(\omega_s/\omega_c)(\omega_c/\omega_0)$ when expressed in terms of the effective coupling constant $\alpha^*_D=-\delta  m\omega_0/2\hbar$ of Eq.~(\ref{alpha*}). The factor $\omega_c/\omega_0$ reflects the fact that the deviation of the system from a strictly 2D geometry, $\omega_0\rightarrow\infty$, is critical for the gate-voltage controlled EDSR. The numerical factor $\omega_s/\omega_c = gm/2m_0$ is about 0.16 for GaSb and InAs and about 0.34 for InSb. Even for a weak magnetic field $\omega_c/\omega_0=0.1$, we estimate $l^\perp_D\sim 5\times 10^{-11}$ cm to $5\times 10^{-10}$\,cm using a typical value $m\sim0.05m_0$ for the mass and also $\delta\approx$ 20\,eV \AA$^3$ for GaAs and 200\,eV \AA$^3$  for  InSb and GaSb.\cite{delta} Hence, under these least favorable conditions the EDSR length $l^\perp_D$ is comparable to the similar length for EPR that is about $l_{\rm EPR}\approx |g|\lambdabar_C/4\approx 10^{-10}$ cm, with $\lambdabar_C=\hbar/m_0c\approx 4\times 10^{-11}$\,cm for the electron Compton length and $|g|\approx 10$ for the electron $g$ factor. 

Remarkably, there are two factors related to the electronic confinement in a quantum well and the experimental geometry that can increase $l^\perp_D$ essentially. The first one is related to the resonance behavior of $l^\perp_D$ in the angle range near $\omega_c\cos\theta\approx |\omega_s|$. The resonance in  $l^\perp_D(\theta)$ is cut off by the level broadening, and the increase in the effective length is about a large factor of $\omega_c/\Gamma\gg1$. The second factor is related to the ratio $\omega_c/\omega_0$ reflecting the deviation of the system from the strict 2D geometry; it is not necessary small. E.g., in the Kato {\it et al.} experiment\cite{KMDGLA03} this ratio was about $\omega_c/\omega_0\approx 0.5$  because a wide parabolic well with an effective width about 50\,nm and a strong magnetic field $B$ = 6T were used. A nearly sixteen time increase of the EDSR intensity due to a four time increase of $\omega_c/\omega_0$ from 0.5 to 2 is seen in Fig.~\ref{fig:BIA}, where the intensity of EDSR for all cases was calculated in arbitrary but the same units based on a fixed confinement energy $\omega_0$. The increase in $\omega_c/\omega_0$ leads to a similar increase in the intensity of the EDSR controlled by Rashba spin-orbit coupling as one can see from Fig.~\ref{fig:Polar}.

The magnitude of the Dresselhaus spin orbit coupling constant $\alpha_D^*$ is controlled by the bulk parameter $\delta$ and cannot be modified significantly. Even in the wide quantum wells used in the Kato {\it et al.} experiments,\cite{KMDGLA03} the effective 2D coupling constant reaches only the value $\alpha^*_D\approx 0.3\times 10^{-10}$\,eV\,cm. This is much less than the typical value of the Rashba constant $\alpha_R\sim 10^{-9}$\,eV\,cm for InAs based quantum wells.\cite{NATE97} Even larger values $\alpha_R\approx (3\div6)\times 10^{-9}$\,eV\,cm were reported in Refs.~[\onlinecite{G00}] and [\onlinecite{alphaR}]. Large $\alpha_R$ values should significantly enhance the efficiency of EDSR whose magnitude can be estimated using Eq.~(\ref{R2D}) and gives the effective magnetic length $l_R^\perp\sim(\alpha_R/\hbar\omega_0)(\omega_s/\omega_c)(\omega_c/\omega_0)$. This suggests that asymmetric quantum wells should have a potential for enhancing EDSR. However, calculations of $\alpha_R$ depend strongly on the boundary conditions,\cite{alpha} and the dependence of $\alpha_R$ on the well width has not been investigated yet. An increase in $\omega_0$ might happen to be a price for increasing $\alpha_R$. From this standpoint, the potential of wurtzite type materials where $\alpha_R$ emerges as a bulk parameter should be investigated in more detail.

The estimates provided above show that electrically manipulating electron spins is preferable to the magnetic one not only because it allows access to electron spins at a nanometer scale but also because  a significantly larger coupling constant can be achieved. In order to reach high efficiency of EDSR in the geometry with an electric field $\tilde{\mbox{\boldmath$E$}}$ perpendicular to the quantum well plane, one needs to use wide asymmetric wells and a strong magnetic field \mbox{\boldmath$B$}  strongly tilted to the quantum well plane. The EDSR should be also stronger in narrow gap semiconductors where the factor $\omega_s/\omega_c$ is typically larger.

Let us  estimate now the advantages of EDSR with an in-plane electric field $\tilde{\mbox{\boldmath$E$}}$. The effective lengths characteristic for the Rashba and Dresselhaus spin-orbit coupling mechanisms are described by Eqs.~(\ref{APLR}) and (\ref{APLD}), respectively. By the order of magnitude, $l_R^\parallel\approx\alpha_R/\hbar\omega_c$ and $l_D^\parallel\approx\alpha_D/\hbar\omega_c$ for $\mbox{\boldmath$B$}\parallel{\hat{\mbox{\boldmath$z$}}}$.  A comparison with the estimates used above for the geometry with a perpendicular to the plane field $\tilde{\mbox{\boldmath$E$}}$ shows that
both small factors in the expressions for the effective lengths do not appear in the equations for the in-plane geometry. The small factor $\omega_c/\omega_0$ has been discussed above and is related to the deviation from the strict 2D limit. This criterion, as well as necessity of using a tilted magnetic field \mbox{\boldmath$B$} are specific for the geometry with a perpendicular to the plane $\tilde{\mbox{\boldmath$E$}}$ and are required to couple this field to in-plane dynamics. With an in-plane electric field the both limitations do not exist any more, and this critical difference manifests itself in Eqs.~(\ref{APLR}) and (\ref{APLD}).

The second small factor, $\omega_s/\omega_0$, originated because of the confinement in the electric field direction; a similar factor appears for EDSR with impurity centers \cite{RS91}. For an in-plane electric field, the motion in the direction of $\tilde{\mbox{\boldmath$E$}}$ is unrestricted and that is why the parameter $\omega_s/\omega_0$ does not appear in Eqs. ~(\ref{APLR}) and (\ref{APLD}).

With typical values of $\alpha_D\approx 0.3\times 10^{-10}$\,eV\,cm  and   $\alpha_R\approx 10^{-9}$ eV cm, $m\approx 0.05m_0$, and $B\approx 1$ T, we get $l_D^\parallel\approx 0.3\times 10^{-6}$ cm and $l_R^\parallel\approx 10^{-5}$ cm correspondingly.  As a result, $l_D^\parallel, l_R^\parallel\gg l_D^\perp, l_R^\perp$ as well as $l_D^\parallel, l_R^\parallel\gg l_{\rm EPR}$, and {\it electrical spin operation by an in-plane electric field should be especially efficient}. Recent experimental data\cite{SLDD05} are in agreement with this prediction of our theory.\cite{APL04} 

The most important quantity characterizing spin operation efficiency by a resonant electric field ${\tilde{\mbox{\boldmath$E$}}}(t)$ is the Rabi frequency $\Omega_R=e{\tilde E}l/\hbar$. With $l\approx \l_R^\parallel, l_D^\parallel\approx 10^{-5}\div10^{-6}$\,cm as estimated above, we find that $\Omega_R\approx 10^{10} {\rm s}^{-1}$ in electric fields as small as only about ${\tilde E}\approx 0.6\div6$\,V/cm.

This extraordinary efficiency of in-plane electric fields bears a promise of operating spins by a set of small vertical gates producing considerable in-plane fields between them. Such a geometry can allow combining the access to electron spins at a nanometer scale typical of vertical gates with the high EDSR intensity typical of in-plane electric fields.  

In conclusion, our theory predicts efficient EDSR with free electrons in quantum wells. Usually, its intensity should be much higher than the intensity of the standard EPR. Electrical spin operation should be extraordinary efficient with an in-plane electric field.  In the perpendicular to the plane electric field geometry, EDSR requires mixing of the in-plane and perpendicular to the plane orbital motions which is provided by a tilted magnetic field. As a result, EDSR in this geometry is significantly weaker than in the in-plane geometry, and semiconductor compounds with large $g$-factors are highly advantageous. The dependence of the EDSR intensity on the magnetic field direction, on both its polar angle and azimuth, is indicative of the competing mechanisms of spin-orbit coupling. 

\section*{Acknowledgments}

We are grateful to T. A. Kennedy for discussion and numerous valuable suggestions. Al.L.E. acknowledges the financial support from DARPA/QuIST program and the Office of the Naval Research. E.I.R. acknowledges financial support from the Harvard Center for Nanoscale Systems and DARPA.

\appendix
\section{Useful identities}
\label{identities}
All formulae for intensities of EDSR, as well as for eigenfrequencies $\omega_\xi$ and $\omega_\eta$, include the auxiliary angle $\gamma$ that should be found from Eq.~(\ref{eqn5}). However, it can be usually eliminated and equations for the intensities of EDSR can be simplified due to the existence of a system of identities that will be derived in this Appendix.

It follows directly from Eqs.~(\ref{eqn6}) that
\be
\omega_\xi^2+\omega_\eta^2=\omega_c^2+\omega_0^2.
\label{eqA1}
\ee
Using Eqs.~(\ref{eqn5}) and (\ref{eqn6}), one arrives at the identities
\be
\omega_\xi^2\omega_\eta^2=\omega_c^2\omega_0^2\cos^2\theta,
\label{eqA2}
\ee
\be
\omega_\xi^2\sin^2\gamma+\omega_\eta^2\cos^2\gamma=\omega_0^2\cos^2\theta,
\label{eqA3}
\ee
\be
\omega_\xi^2\cos^2(\theta+\gamma)+\omega_\eta^2\sin^2(\theta+\gamma)=\omega_c^2\cos^2\theta.
\label{eqA4}
\ee
>From Eqs.~(\ref{eqA2}) and (\ref{eqA3}) follows the identity
\be
{{\sin^2\gamma}\over{\omega_\eta^2}}+{{\cos^2\gamma}\over{\omega_\xi^2}}={1\over{\omega_c^2}}
\label{eqAa}
\ee
that ensures fulfillment of the commutation relations $[{\hat k}_x,{\hat x}]=[{\hat k}_y,{\hat y}]=-i$. Subtracting Eqs.~(\ref{eqA3}) and (\ref{eqA4}) from (\ref{eqA1}), one finds
\be
\omega_\xi^2\cos^2\gamma+\omega_\eta^2\sin^2\gamma=\omega_c^2+\omega_0^2\sin^2\theta,
\label{eqA5}
\ee
\be
\omega_\xi^2\sin^2(\theta+\gamma)+\omega_\eta^2\cos^2(\theta+\gamma)=\omega_c^2\sin^2\theta+\omega_0^2.
\label{eqA6}
\ee
Using Eq.~(\ref{eqA6}), one immediately finds
\be
{\cos^2(\theta+\gamma)\over \omega_\xi^2-\omega_s^2}+{\sin^2(\theta+\gamma)\over \omega_\eta^2-\omega_s^2}={\omega_0^2+\omega_c^2\sin^2\theta-\omega_s^2\over(\omega_\xi^2-\omega_s^2)(\omega_\eta^2-\omega_s^2)}\,,
\label{eqA7}
\ee
and applying Eqs.~(\ref{eqA2}) and (\ref{eqA4}), one arrives at
\be
{\cos^2(\theta+\gamma)\omega_\xi^2\over \omega_\xi^2-\omega_s^2}+{\sin^2(\theta+\gamma)\omega_\eta^2\over \omega_\eta^2-\omega_s^2}={\omega_c^2\cos^2\theta(\omega_0^2-\omega_s^2)\over(\omega_\xi^2-\omega_s^2)(\omega_\eta^2-\omega_s^2)}\,.
\label{eqA8}
\ee
Two similar equations can be written for the angle $\gamma$. Remarkably, the right hand sides of Eqs.~(\ref{eqA3}) - (\ref{eqA6}) do not depend on $\gamma$, and the same is true for the right hand sides of Eqs.~(\ref{eqA7}) and (\ref{eqA8}) because
\be
(\omega_\xi^2-\omega_s^2)(\omega_\eta^2-\omega_s^2)=\omega_0^2\omega_c^2\cos^2\theta-\omega_s^2(\omega_0^2+\omega_c^2-\omega_s^2),
\label{eqA9}
\ee
as follows from Eqs.~(\ref{eqA1}) and (\ref{eqA2}).

>From Eqs.~(\ref{eqn5}) and (\ref{eqn6}) one can also derive an identity
\be
\omega_\xi^2\tan\gamma=\omega_\eta^2\tan(\theta+\gamma)~,
\label{eqA10}
\ee
from which the following relations follow
\bea
{{\sin(\theta+\gamma)\cos\gamma}\over{\omega_\xi^2}}&=&{{\sin\gamma\cos(\theta+\gamma)}\over{\omega_\eta^2}}={{\sin2\gamma}\over{2\omega_0^2\cos\theta}}\nonumber\\
&=&{{\sin2(\theta+\gamma)}\over{2\omega_c^2\cos\theta}}.
\label{eqA11}
\eea
In turn, a set of new identities follows from Eq.~(\ref{eqA11})
\bea
\cos\gamma&=&{{\omega_\xi^2}\over{\omega_c^2}}{{\cos(\theta+\gamma)}\over{\cos\theta}},\,\,
\cos(\theta+\gamma)={{\omega_\eta^2}\over{\omega_0^2}}{{\cos\gamma}\over{\cos\theta}}\,,\nonumber\\
\sin\gamma&=&{{\omega_\eta^2}\over{\omega_c^2}}{{\sin(\theta+\gamma)}\over{\cos\theta}},\,\,
\sin(\theta+\gamma)={{\omega_\xi^2}\over{\omega_0^2}}{{\sin\gamma}\over{\cos\theta}}\,.\nonumber\\
\label{eqA12}
\eea
These identities, together with Eq.~(\ref{commrel}), allow proving the commutation relation $[{\hat x},{\hat y}]=0.$ Also, using Eq.~(\ref{eqn5}), one easily finds
\bea
\tan2\gamma&=&{{\sin2\theta}\over{(\omega_c/\omega_0)^2-\cos2\theta}}~,\nonumber\\
\tan2(\theta+\gamma)&=&{{\sin2\theta}\over{\cos2\theta-(\omega_0/\omega_c)^2}}~.
\label{eqA13}
\eea
These equations allow one to derive the identity
\be
\omega_\xi^2-\omega_\eta^2={\omega_0^2\sin2\theta\over \sin2\gamma}={\omega_c^2\sin2\theta\over \sin2(\theta+\gamma)}~.
\label{eqA14}
\ee

All the above identities have been derived directly from Eqs.~(\ref{eqn5}) and (\ref{eqn6}) and are valid for an arbitrary ratio $\omega_c/\omega_0$ and an arbitrary value of $\theta$, $0\leq\theta\leq\pi$.

Finally, using Eqs.~(\ref{eqA13}), one can derive explicit expressions for $\sin2\gamma$ and $\sin2(\theta+\gamma)$ that allow one to find $\omega_\xi$ and $\omega_\eta$ from Eqs.~(\ref{eqA1}) and (\ref{eqA14}). For $0\leq\theta\leq\pi/2$,
\bea
\sin2\gamma&=&-{{\omega_0^2}\over{\Omega^2}}\sin2\theta~{\rm sign}\{\omega_0-\omega_c\}~,\nonumber\\
\sin2(\theta+\gamma)&=&-{{\omega_c^2}\over{\Omega^2}}\sin2\theta~{\rm sign}\{\omega_0-\omega_c\}~,
\label{eqA15}
\eea
with $\Omega^2(\theta)$ defined by Eq.~(\ref{Omega}). Equation~(\ref{eqA13}) defines $\gamma$ with the accuracy to the phase $\pm\pi/2$. The choice of the phase made in Eq.~(\ref{eqA15}) results in the energy spectrum of Eq.~(\ref{omsexpl}) shown in Fig.~\ref{fig:spectrum}. With this choice of the phase, Eq.~(\ref{eqA14}) takes the form
\be
\omega_\eta^2-\omega_\xi^2=\Omega^2(\theta){\rm sign}\{\omega_0-\omega_c\}.
\label{omegasquar}
\ee
It indicates that the difference $\omega_\xi-\omega_\eta$, when considered as a function of $\omega_c/\omega_0$, changes sign at $\omega_c=\omega_0$.

\section{Several sums contributing to EDSR}
\label{coeffb}

For the calculation of $C_{j;\zeta\zeta'}$ we need to find bilinear combinations of the coefficients $X_{\zeta'}$, $Y_{\zeta'}$, and $Z_{\zeta'}$ appearing in the brackets of Eq. \ref{Ds}. Straightforward calculation of these coefficients based on their definition by Eq.~(\ref{eqn15}) results in
\bea
|X_\zeta|^2-|Y_\zeta|^2&=&\beta_\zeta b_{X-Y},
X_\zeta Y_\zeta^*+Y_\zeta X_\zeta^*=\beta_\zeta b_{XY},\nonumber\\
 {\rm and}~~~b_{YX}&=&b_{XY},
\label{bs}
\eea
where $\beta_{\xi}=-m\omega_\xi/2\hbar$, $\beta_{\eta}=m\omega_\eta/2\hbar$, and all other combinations can be found from Eq.~(\ref{bs}) by cyclic transpositions of $X$, $Y$, and $Z$. Here
\bea
b_{X-Y}&=&\frac{1}{2}\sin2(\theta+\gamma)\tan\theta\cos2\phi~,\nonumber\\
b_{Y-Z}&=&-[\cos2(\theta+\gamma)+{1\over 2}\sin2(\theta+\gamma)\tan\theta\cos^2\phi]~,\nonumber\\
b_{Z-X}&=&[\cos2(\theta+\gamma)+{1\over2}\sin2(\theta+\gamma)\tan\theta\sin^2\phi]~,\nonumber\\
b_{XY}&=&\frac{1}{2}\sin2(\theta+\gamma)\tan\theta\sin2\phi~,\nonumber\\
b_{YZ}&=&\sin2(\theta+\gamma)\sin\phi~,\nonumber\\
b_{ZX}&=&\sin2(\theta+\gamma)\cos\phi~.
\label{bms}
\eea
Sums standing in the numerators of Eq.~(\ref{z2}) involve the expressions
\bea
R_{j\xi}+R_{j\xi}^*&=&2\omega_c\sqrt{m/2\hbar\omega_\xi}\cos\gamma~ G_j^+, \nonumber\\
R_{j\eta}+R_{j\eta}^*&=&2\omega_c\sqrt{m/2\hbar\omega_\eta}\sin\gamma~ G_j^+,
\label{sum}
\eea
and differences standing in the same numerators involve the expressions
\bea
R_{x\zeta}-R_{x\zeta}^*&=&-2i\sqrt{m\omega_\zeta/2\hbar}~f_\zeta G_x^-~, \nonumber\\
R_{y\zeta}-R_{y\zeta}^*&=&-2i\sqrt{m\omega_\zeta/2\hbar}~f_\zeta G_y^-~, \nonumber\\
R_{z\zeta}-R_{z\zeta}^*&=&(-)^{m_\zeta}2i\sqrt{m\omega_\zeta/2\hbar}~f_{\bar\zeta}~,
\label{dif}
\eea
where
\be
\mbox{\boldmath$G$}^+=(-\sin\phi,\cos\phi,0),\,\,\mbox{\boldmath$G$}^-=(\cos\phi,\sin\phi,0),
\label{Gs}
\ee
$\bar \zeta$ is defined as ${\bar\xi}=\eta$ and ${\bar\eta}=\xi$, and $m_\zeta$ is defined as $m_\xi=0,m_\eta=1$.

Using Eqs.~(\ref{sum}) and (\ref{dif}), it is convenient to rewrite two groups of coefficients needed for calculating numerators of Eqs.~(\ref{z2}) as
\be
l_{z\zeta}\omega_\zeta(C_{j;\zeta,\zeta'}+C_{j;\zeta,\zeta'}^*)=-(-1)^{m_{\zeta}}{\omega_\zeta^2\sin2(\theta+\gamma)\over 2\omega_c\cos\theta}\beta_{\zeta'}M^+_j,\nonumber\\
\ee
\bea
l_{z\zeta}(C_{j;\zeta\zeta'}-C_{j;\zeta\zeta'}^*)&=&i\beta_{\zeta'}\left[\frac{1}{2}(-1)^{m_\zeta}\sin2(\theta+\gamma)M^-_j\right.\nonumber\\
&+&\left.(1-f_\zeta^2)N_j\right],
\eea
where $\mbox{\boldmath$M$}^\pm$ and $\mbox{\boldmath$N$}$ are 3D vectors 
\bea
M_j^\pm&=&G_j^\pm d_j+(\mbox{\boldmath$G$}^\pm\times\mbox{\boldmath$D$})_j~,\nonumber\\
\mbox{\boldmath$N$}&=&(b_{ZX},-b_{ZY},-b_{X-Y})~.
\label{mpmN}
\eea
Using these expressions, one can transform Eq.~(\ref{z2}) by collecting terms having common factors $2n_\zeta+1$. Then the following sums emerge
\bea
l_{z\zeta}\omega_\zeta{C_{j;\zeta\zeta}+C_{j;\zeta\zeta}^*\over \omega_\zeta^2-\omega_s^2}&+&l_{z{\bar\zeta}}\omega_{\bar\zeta}{C_{j;{\bar\zeta}\zeta}+C_{j;{\bar\zeta}\zeta}^*\over \omega_{\bar\zeta}^2-\omega_s^2}\nonumber\\
&=& M^+_{j}{\beta_\zeta\omega_c\omega_s^2\sin\theta\over {\cal D}(\theta)}~,
\label{comb1}
\eea
\bea
l_{z\zeta}{C_{j;\zeta\zeta}-C_{j;\zeta\zeta}^*\over \omega_\zeta^2-\omega_s^2}&+&l_{z{\bar\zeta}}{C_{j;{\bar\zeta}\zeta}-C_{j;{\bar\zeta}\zeta}^*\over \omega_{\bar\zeta}^2-\omega_s^2}\nonumber\\
=-{i\over 2}M_{j}^-{\beta_\zeta\omega_c^2\sin2\theta\over {\cal D}(\theta)}&+&iN_{j}{\beta_\zeta(\omega_c^2\cos^2\theta-\omega_s^2)\over {\cal D}(\theta)}~.
\label{comb2}\nonumber\\
\eea
Plugging these equations into Eq.~(\ref{3DTR}), we arrive at
\bea
&&\langle n_\xi,n_\eta,\uparrow\vert\hat{z}_{\rm so}\vert n_\xi,n_\eta,\downarrow\rangle\nonumber\\
&=&-{\delta\over \hbar}\sum_{\zeta}(2n_\zeta+1)\beta_\zeta\left[{\cal M}^+{\omega_c\omega_s^2\sin\theta\over {\cal D}(\theta)}\right.\nonumber\\
&-&\left.\frac{i}{2}{\cal M}^-{\omega_s\omega_c^2\sin2\theta\over {\cal D}(\theta)}+i{\cal N}{\omega_s(\omega_c^2\cos^2\theta-\omega_s^2)\over {\cal D}(\theta)}\right]~,\nonumber\\
\label{z3}
\eea
with
\[{\cal M}^\pm=(U_{jx'}-iU_{jy'})M_j^\pm,\,\, {\cal N}=(U_{jx'}-iU_{jy'})N_j.\]
Calculating these coefficients by using Eqs.~(\ref{eqn2}) and (\ref{bms}), we find
\bea
&&{\cal M}^+\nonumber\\
&=&\sin2\phi[(7/4)\sin2(\theta+\gamma)\sin\theta+\cos2(\theta+\gamma)\cos\theta]\nonumber\\
&-&i\cos2\phi\cos2(\theta+\gamma)~,\nonumber\\
&&{\cal M}^-\nonumber\\
&=&-\cos2\phi[\cos\theta\cos2(\theta+\gamma)+{3\over2}\sin2(\theta+\gamma)\sin\theta]\nonumber\\
&+&i\sin2\phi[{1\over4}\sin2(\theta+\gamma)\tan\theta-\cos2(\theta+\gamma)]~,\nonumber\\
&&{\cal N}\nonumber\\
&=&\sin2(\theta+\gamma)[\cos2\phi(\cos\theta+{1\over2}\sin\theta\tan\theta)+i\sin2\phi]~.\nonumber\\
\label{MMN}
\eea
From these equations follows the final equation (\ref{Answer}).

\newpage

\begin{figure}
\caption{(Color online) The energy spectrum of electrons confined in a parabolic quantum well in a tilted magnetic field. Here, $\omega_0$ is the confinement frequency, $\omega_c=eB/mc$ is the cyclotron frequency, $\theta$ is the polar angle of the field \mbox{\boldmath$B$}, $\omega_\xi$ and $\omega_\eta$ are the frequencies of two eigenmodes $\omega_\zeta$, $\zeta=\xi,\eta$. (a) The dependence of $\omega_\zeta$ on the ratio $\omega_c/\omega_0$ for three values of $\theta$. Cyclotron-like modes are shown by dotted lines and confinement-like modes by full lines. For $\theta=0$ these modes intersect, while for $\theta\neq0$ they interchange at $\omega_c=\omega_0$. (b) The dependence of $\omega_\zeta$ on $\theta$ for three values of $\omega_c/\omega_0$. The modes retain their identity in the whole interval $0\leq\theta\leq\pi/2$, and $\omega_\eta>\omega_\xi$ for $\omega_c<\omega_0$ while $\omega_\eta<\omega_\xi$ for $\omega_c>\omega_0$. At the degeneracy point, $\omega_c=\omega_0$, the two $\omega_{\eta,\xi}$ branches have no specific identity. They are separatrices dividing the plane into one $\eta$ (gray) and two $\xi$ (white) regions.}
\label{fig:spectrum}
\end{figure} 

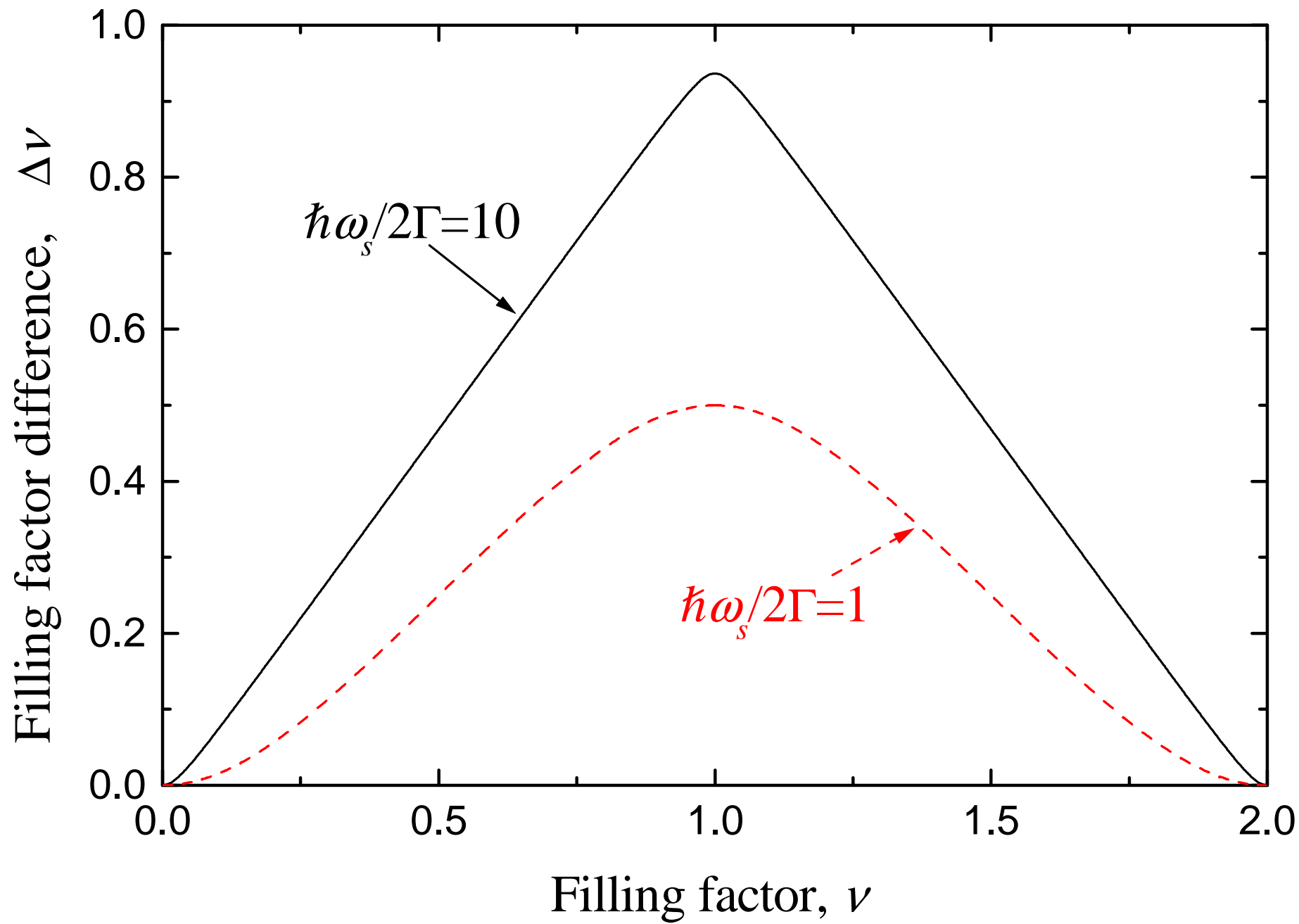
\begin{figure}
\caption{(Color online) The difference of the filling factors, $\Delta\nu$, of the two components of a spin doublet of an isolated energy level as a function of the total filling factor of this level $0\leq\nu\leq2$. For a small level width $\Gamma$, its shape is close to a triangular form, and an up-cusp develops at $\nu=1$. In a plot including adjacent energy levels, down-cusps at the end of this interval will arise. With increasing $\Gamma$, the cusps are smeared. However, at $\Gamma$ as large as $\Gamma=\hbar\omega_s/2$, pronounced extrema at $\nu=1$ and $\nu=0,2$ are still seen.}
\label{fig:TwoLevel}
\end{figure}

\begin{figure}
\caption{(Color online) The dependence of the EDSR intensity on the polar angle $\theta$ of a magnetic field \mbox{\boldmath$B$} for Rashba spin-orbit coupling, Eq.~(\ref{eqR}). The electric field is perpendicular to the confinement plane, ${\tilde{\mbox{\boldmath$E$}}}\parallel{\hat{\bf z}}$. The ratio of spin-flip and cyclotron frequencies, $\omega_s/\omega_c=-0.17$, is typical of InAs; $\omega_0$ being the confinement frequency. The calculations were performed for three values of the level width $\Gamma$, as specified in the figure. For $\mbox{\boldmath$B$}\parallel{\hat{\bf z}}$, electrons completely populate the ground level, $n_\xi=n_\eta=0$. (a) Strong confinement, $\omega_c/\omega_0=0.5$. With increasing $\theta$ electrons populate higher cyclotron-like $\xi$ levels, while they remain at the ground $\eta$ level. (b) Weak confinement, $\omega_c/\omega_0=2$. With increasing $\theta$ electrons populate higher $\eta$ levels, while they remain at the ground $\xi$ level. In both cases, the principal peak originates from the resonance $\omega_\zeta=|\omega_s|$ with $\zeta=\xi$ for (a) and $\zeta=\eta$ for (b). The fine structure of the spectrum, distinctly seen for small level width $\Gamma=0.05\omega_0$, originates from populating higher $n_\zeta$ levels in a tilted magnetic field \mbox{\boldmath$B$}. Up-cusps are marked by up-arrows and down-cusps by down-arrows; for details see text. In (b), the inflexion point seen between the $\nu=5$ cusp and the EDSR maximum reflects an abrupt change in the population of intersecting levels at $\omega_\eta=|\omega_s|$. Inserts show the dependence of the chemical potential $\eta(\nu)$ on the filling factor $\nu$ and the quantum numbers of partially populated states, $n_\xi$ or $n_\eta$, that contribute to spin-flip transitions inside the various regions of filling factor values. }
\label{fig:Polar}
\end{figure}

\begin{figure}
\caption{The dependence of the EDSR intensity on the magnetic field direction for a (001) quantum well. The electric field is perpendicular to the QW plane, ${\tilde{\mbox{\boldmath$E$}}}\parallel{\hat{\bf z}}$. The electron concentration is low , hence, for all $\theta$ values electrons populate only the ground level, $n_\xi,n_\eta=0$. The level width, $\Gamma=0.2\omega_0$; $\omega_c/\omega_0=0.5$, $\omega_s/\omega_c=-0.17$.  (a) 2D Dresselhaus spin-orbit coupling mechanism, Eq.~(\ref{eqDr}). The four-fold symmetry of the figure reflects the symmetry of the square of the spin-flip transition matrix element. (b) Interference of 2D Dresselhaus and Rashba  spin-orbit coupling mechanisms. The two-fold symmetry of the figure reflects the $\mbox{\boldmath$C$}_{2v}$ symmetry of the Hamiltonian ${\hat H}_D$; coupling constants have been chosen as $\alpha_D=\alpha_R$. }
\label{fig:AzimuthPerp}
\end{figure}

\begin{figure}
\caption{The dependence of the EDSR intensity on the magnetic field direction for a (001) quantum well. The electric field is in the QW plane, ${\tilde{\mbox{\boldmath$E$}}}\perp{\hat{\bf z}}$. The electron concentration is low; parameter values are the same as in Fig.~\ref{fig:AzimuthPerp}. (a) and (b) -- Rashba spin-orbit coupling, Eq.~(\ref{eqR}). For (a,) the electric field ${\tilde{\mbox{\boldmath$E$}}}$ is parallel to the projection ${\mbox{\boldmath$B$}}_\perp$ of the field $\mbox{\boldmath$B$}$ onto the confinement plane, ${\tilde{\mbox{\boldmath$E$}}}\parallel{\mbox{\boldmath$B$}}_\perp$, hence, the intensity is rotationally symmetric. For (b), the electric field is parallel to the (110) crystallographic axis, ${\tilde{\mbox{\boldmath$E$}}}\parallel(110)$. (c) and (d) -- 2D Dresselhaus spin-orbit coupling, Eq.~(\ref{eqDr}). For (c), the electric field ${\tilde{\mbox{\boldmath$E$}}}$ is parallel to $\mbox{\boldmath$B$}_\perp$, ${\tilde{\mbox{\boldmath$E$}}}\parallel{\mbox{\boldmath$B$}}_\perp$, hence, the intensity shows four-fold symmetry. For (d),  ${\tilde{\mbox{\boldmath$E$}}}\parallel(110)$, and the gross features are similar to those of figure (b). For (e), ${\tilde{\mbox{\boldmath$E$}}}\parallel{\mbox{\boldmath$B$}}_\perp$, the two spin-orbit coupling mechanisms with $\alpha_D=\alpha_R$ interfere; the symmetry is lowered because of their interference.}
\label{fig:InPlane}
\end{figure}

\begin{figure}
\caption{(Color online) The dependence of the EDSR intensity on the polar angle $\theta$ of a magnetic field \mbox{\boldmath$B$} for Dresselhaus spin-orbit coupling, Eq.~(\ref{eq11}), for two values of the azimuth $\phi$. The electric field is perpendicular to the QW plane, ${\tilde{\mbox{\boldmath$E$}}}\parallel{\hat{\bf z}}$. The notations and basic parameter values are the same as in Fig.~\ref{fig:Polar}. Similarly to Fig.~\ref{fig:Polar}, the spectra show the principal maximum at $\omega_\zeta=|\omega_s|$ and the fine structure of up- and down-cusps indicating repopulation of different energy levels with $\theta$ changing. The major distinctions from Fig.~\ref{fig:Polar} include (i) a minimum at $\theta=0.42\times\pi/2$ originating from a zero in the effective coupling constant $\alpha^{\rm eff}_D(n_\xi,n_\eta)$ of Eq.~(\ref{alpha*}), (ii) additional minima in panels c and d originating from the zeros of functions $F_c(\theta)$ and $F_s(\theta)$ at $\theta=0.82\times\pi/2$ and $\theta=0.61\times\pi/2$, respectively, and (iii) smearing of the main peak in panel (a) because of the slow angular dependence of the envelope function shown in the inset to that panel. For details see the text. The inserts to panels (a) and (b) demonstrate that the results found for the strong confinement limit still retain reasonable accuracy for $\omega_c=\omega_0/2$. The inserts to panels (c) and (d) demonstrate the positions of the zeros of functions $F_c$ and $F_s$.}
\label{fig:BIA}
\end{figure}






\begin{thebibliography}{99}
\bibitem{g-corr} Spin-orbit corrections to the energy levels and $g$-factor have been recently calculated both for parabolic quantum wells\cite{PZ05} and parabolic quantum dots.\cite{PC05}
\bibitem{error} Equation (\ref{R2D}) corrects the corresponding equation of Ref.~\onlinecite{RE03} where in the parentheses $(\omega_c+\omega_s)$ there was the opposite sign at $\omega_s$.
\bibitem{semimag} This argument holds when $\omega_s\alt\omega_0,\omega_c$, that is typical for regular semiconductors. The opposite inequality, $\omega_s\gg\omega_c$, is typical of semimagnetic semiconductors, but we do not discuss them because an additional mechanism of SO coupling operates there.\cite{KRS}
\bibitem{zeros} We note that the phenomenological width $\Gamma$ has been incorporated only into the function ${\cal D}(\theta)$ to cut-off the singularity at $\omega_\zeta=|\omega_s|$ and into the calculations of the Fermi level position $\eta(\theta)$. Adding the same width $\Gamma$ to the cyclotron frequency in the numerators of $F_c$ and $F_s$ would transform their zeros into deep minima in $|F_c(\theta)|^2$ and $|F_s(\theta)|^2$.
\end{thebibliography}
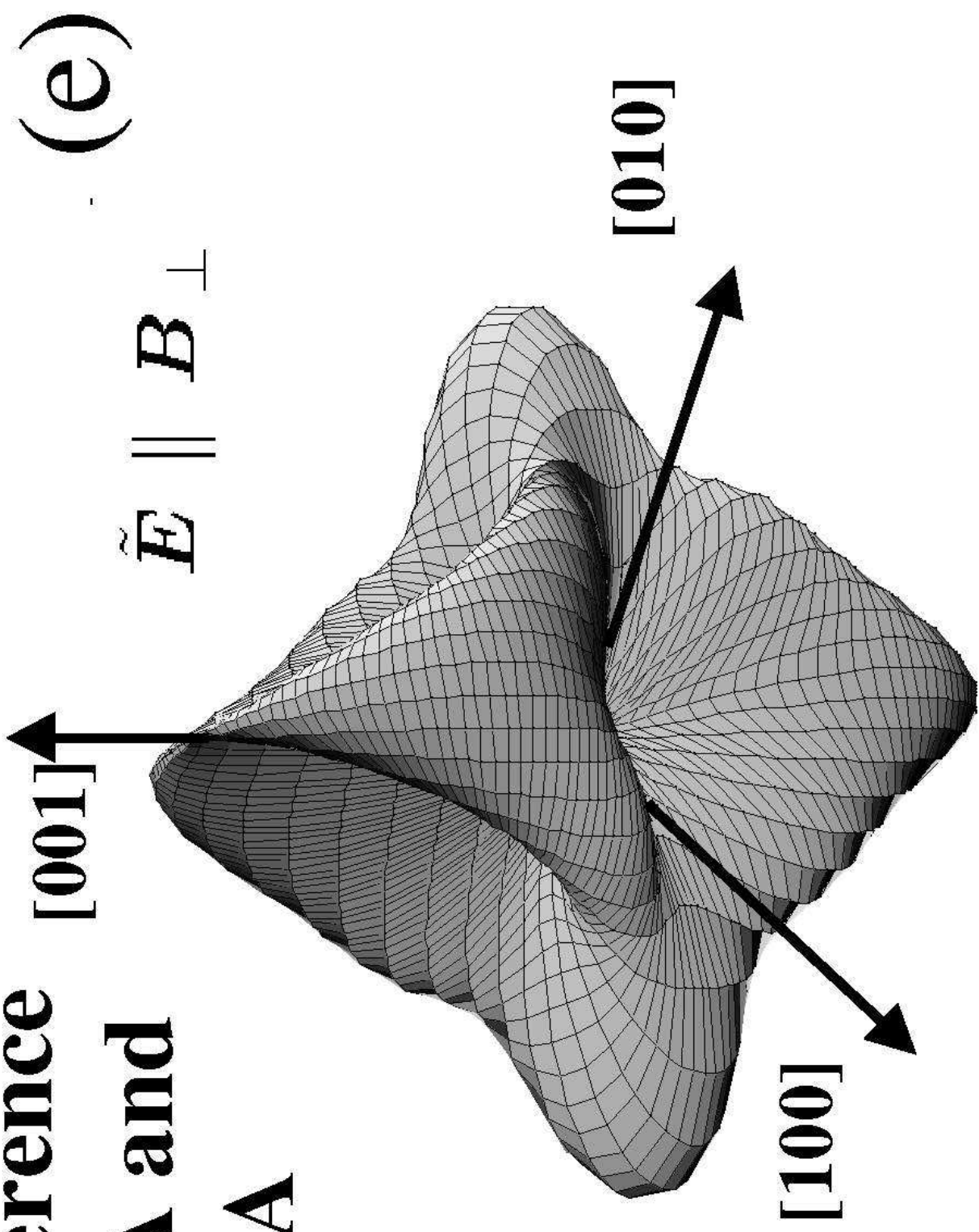
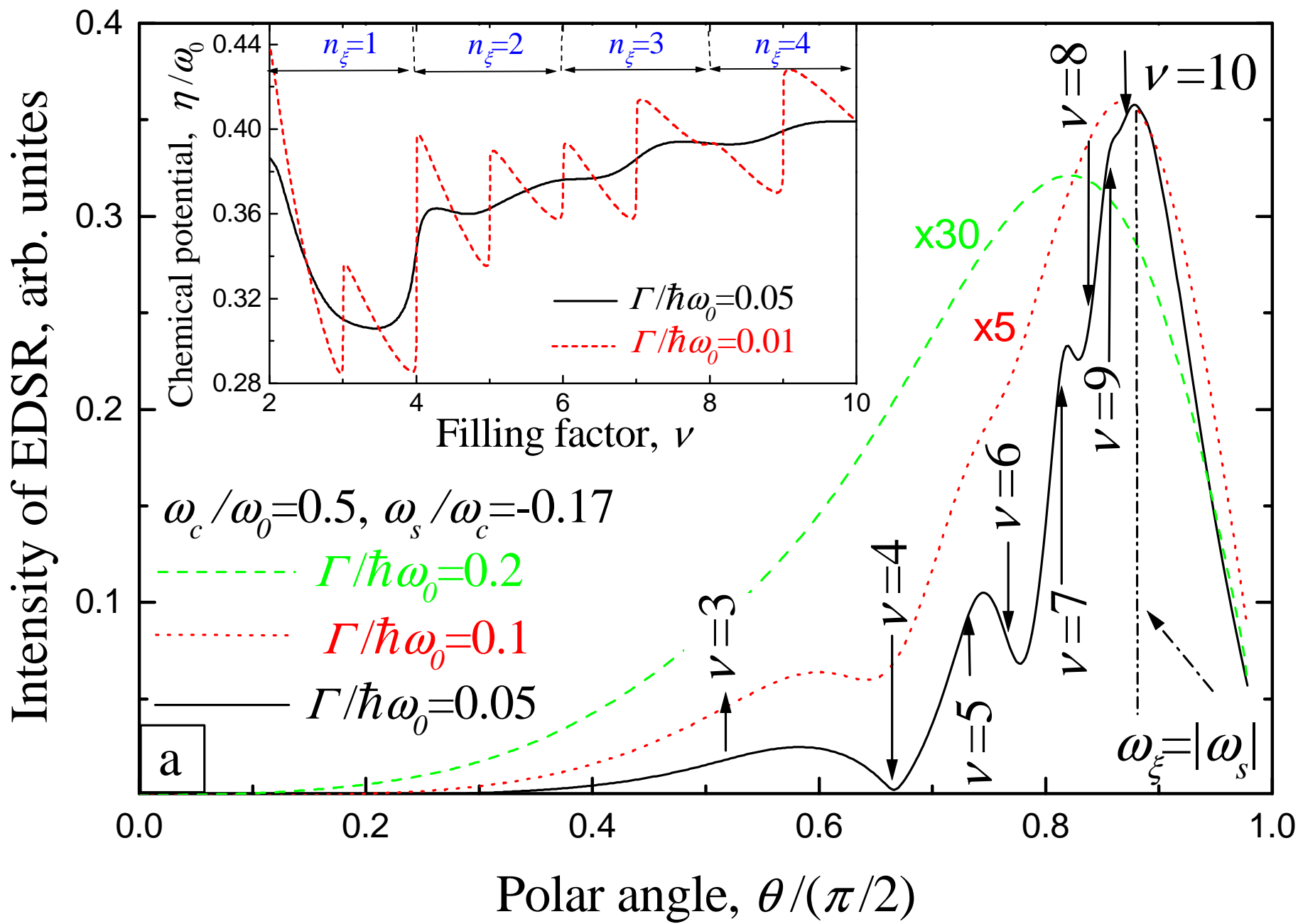
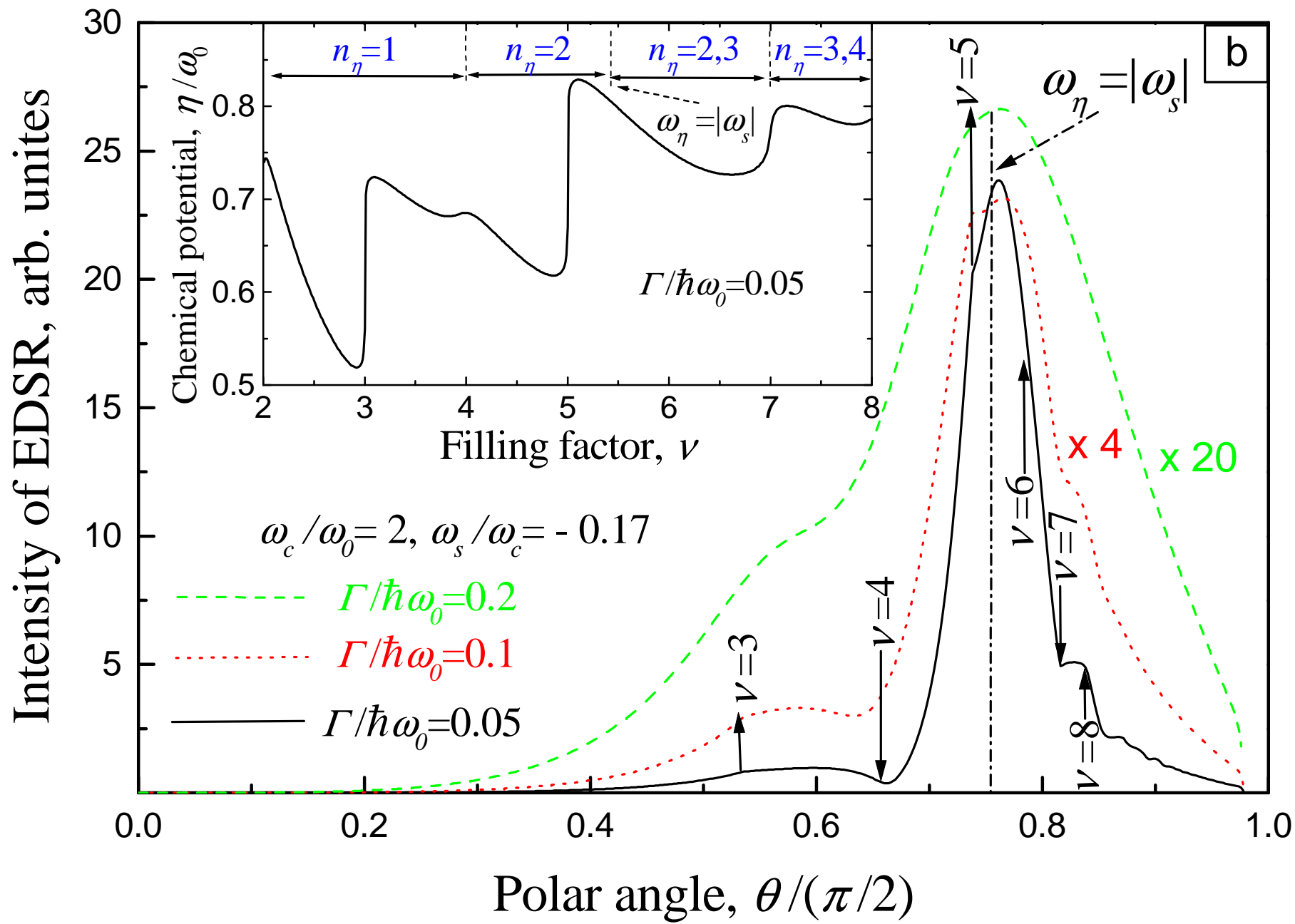
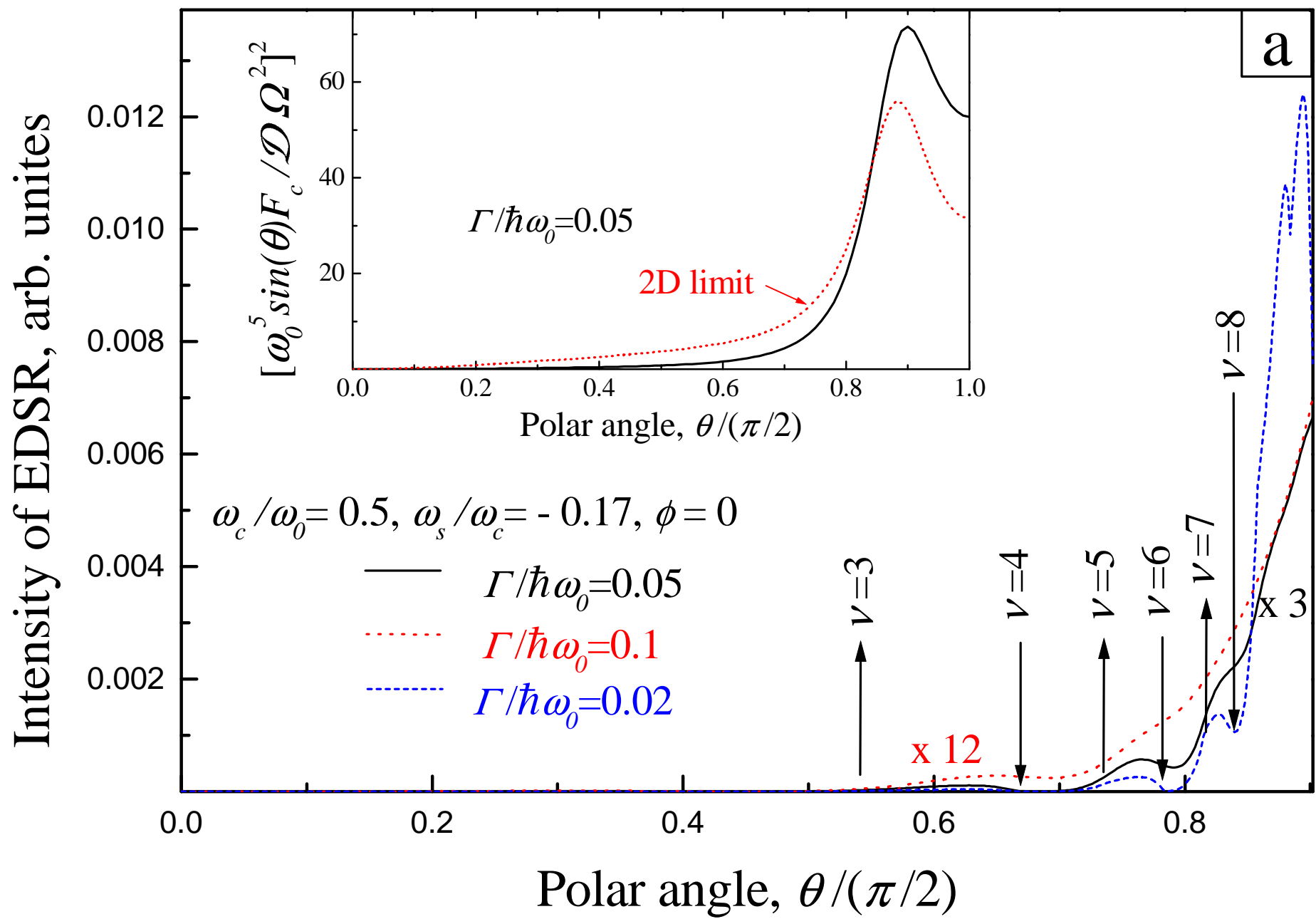
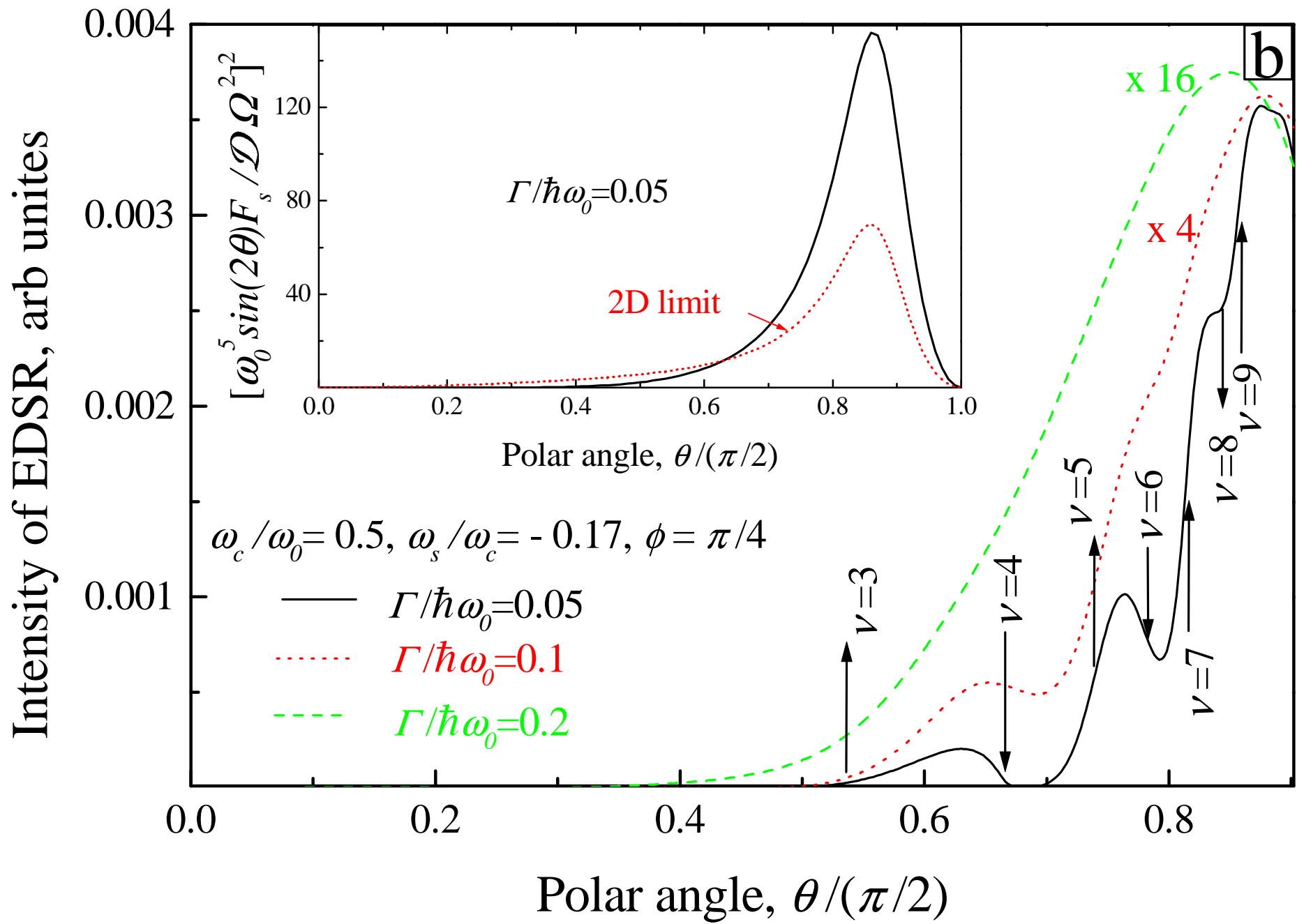
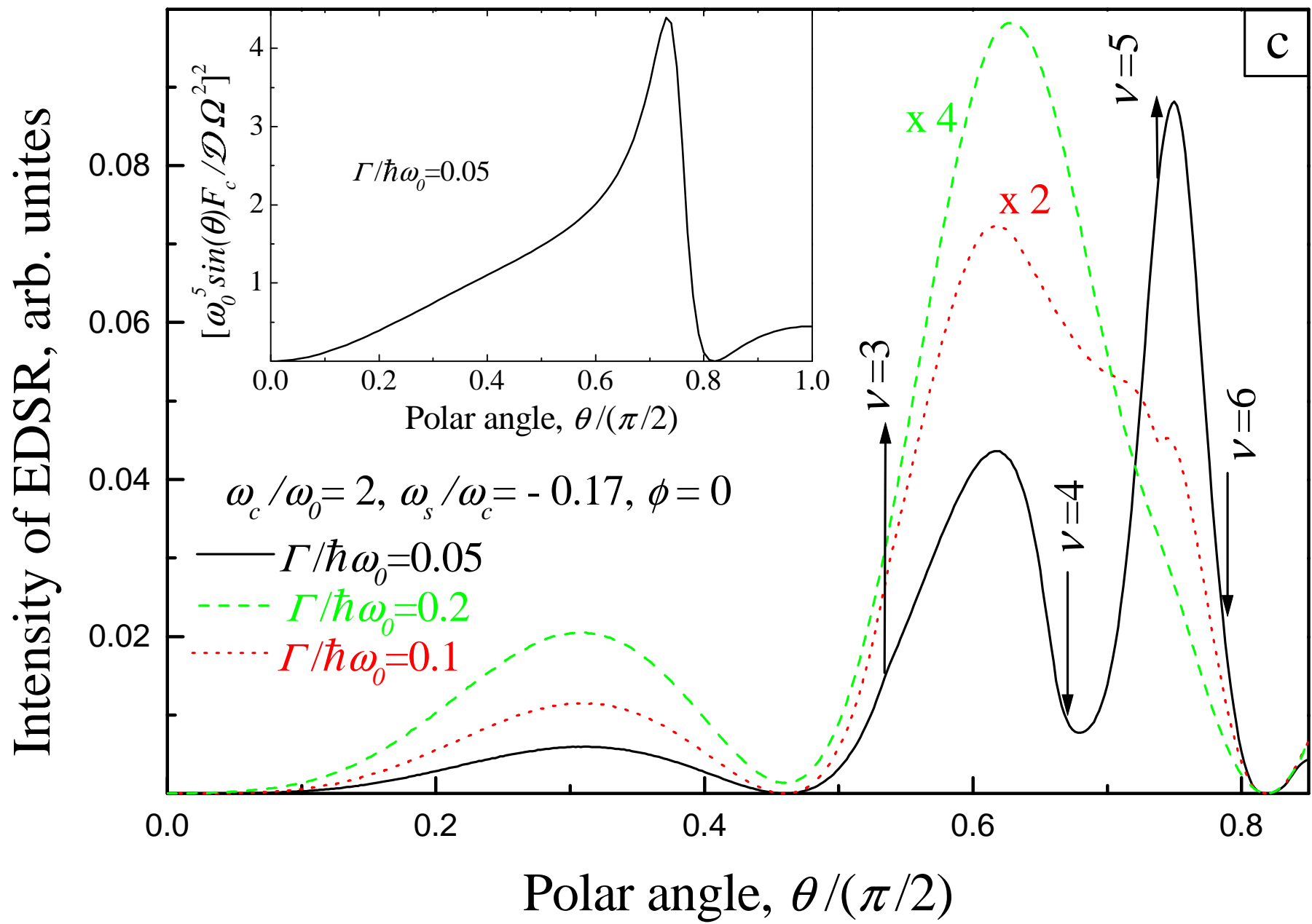
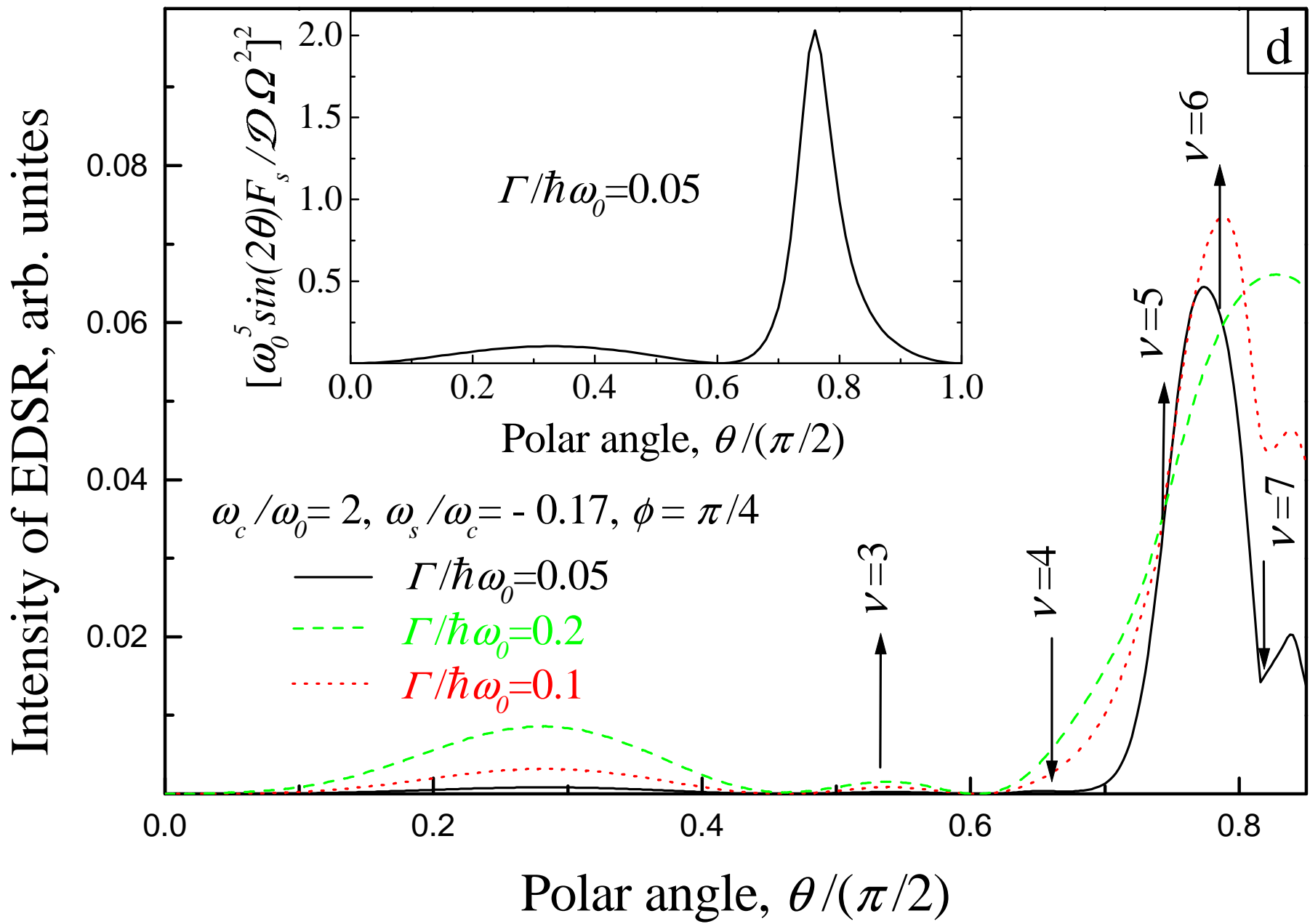
\end{document}